\documentclass[%
 reprint,
 superscriptaddress,
 amsmath,amssymb,
 aps,
prl
]{revtex4-2}

\usepackage{amsmath}
\usepackage{amsthm}
\usepackage[colorlinks=true,urlcolor=blue,citecolor=blue,linkcolor=blue]{hyperref}
\usepackage[shortlabels]{enumitem}
\usepackage{amsfonts}
\usepackage{graphicx}
\usepackage{bbold}
\usepackage{color}
\usepackage{hyperref}
\usepackage{verbatim}
\usepackage{cases}
\usepackage{amsfonts}
\usepackage{amssymb}
\usepackage[]{mathrsfs}
\usepackage{xcolor}
\usepackage{epsfig}
\usepackage{bm}
\usepackage{setspace}
\usepackage{enumerate}
\usepackage{dcolumn}
\usepackage{bm}
\usepackage{enumitem}

\usepackage{graphicx}
\usepackage{dcolumn}
\usepackage{bm}
\usepackage[FIGTOPCAP,small,tight]{subfigure}
\usepackage{color}

\newcommand{\pt}{\mathcal{PT}}

\begin{document}
\preprint{APS/123-QED}

\title{On-demand Parity-Time symmetry in a lone oscillator through complex, synthetic gauge fields}

\author{Mario A. Quiroz-Ju\'{a}rez}
\affiliation{Departamento de F\'{i}sica, Universidad Aut\'onoma
Metropolitana Unidad Iztapalapa, San Rafael Atlixco 186, 09340 Cd. Mx., M\'exico}

\author{Kaustubh S. Agarwal}
\affiliation{Department of Physics, Indiana University - Purdue University Indianapolis (IUPUI), Indianapolis, Indiana 46202 USA}

\author{Zachary A. Cochran}
\affiliation{Department of Physics, Indiana University - Purdue University Indianapolis (IUPUI), Indianapolis, Indiana 46202 USA}

\author{Jos\'{e} L. Arag\'{o}n}
\affiliation{Centro de F\'{i}sica Aplicada y Tecnolog\'{i}a Avanzada, Universidad Nacional Aut\'onoma de M\'exico, Boulevard Juriquilla 3001, 76230 Quer\'{e}taro, M\'exico}

\author{Yogesh N. Joglekar}
\email{yojoglek@iupui.edu}
\affiliation{Department of Physics, Indiana University - Purdue University Indianapolis (IUPUI), Indianapolis, Indiana 46202 USA}

\author{Roberto de J. Le\'on-Montiel}
\email{roberto.leon@nucleares.unam.mx}
\affiliation{Instituto de Ciencias Nucleares, Universidad Nacional Aut\'onoma de M\'exico, Apartado Postal 70-543, 04510 Cd. Mx., M\'exico}

\date{\today}

\begin{abstract}
What is the fate of an oscillator when its inductance and capacitance are varied while its frequency is kept constant? Inspired by this question, we propose a protocol to implement parity-time ($\mathcal{PT}$) symmetry in a lone oscillator. Different forms of constrained variations lead to static, periodic, or arbitrary balanced gain and loss profiles, that can be interpreted as purely imaginary gauge fields. With a state-of-the-art, dynamically tunable $LC$ oscillator comprising synthetic circuit elements, we demonstrate static and Floquet $\mathcal{PT}$ breaking transitions, including those at vanishingly small gain and loss, by tracking the circuit energy.  Concurrently, we derive and observe conserved quantities in this open, balanced gain-loss system, both in the static and Floquet cases. Lastly, by measuring the circuit energy, we unveil a giant dynamical asymmetry along exceptional point (EP) contours that emerge symmetrically from the Hermitian degeneracies at Floquet resonances. Distinct from material or parametric gain and loss mechanisms, our protocol enables on-demand parity-time symmetry in a minimal classical system--- a single oscillator--- and may be ported to other realizations including metamaterials and optomechanical systems.
\end{abstract}

\maketitle



\noindent{\bf Introduction.} Non-Hermitian Hamiltonians that are invariant under combined operations of parity and time-reversal (or more generally, an antilinear symmetry) are called $\mathcal{PT}$ symmetric. Since their discovery in the late 1990s~\cite{Bender1998,Bender2001} and first experimental realization in coupled optical waveguides a decade later~\cite{Duchesne2009,Ruter2010}, the field of $\mathcal{PT}$-symmetric systems has grown increasingly diverse~\cite{Feng2017,ElGanainy2018,Gupta2019}. This expansive growth is driven by the realization that a $\mathcal{PT}$-symmetric Hamiltonian represents an open system with balanced, but separated, gain and loss. A $\mathcal{PT}$-symmetric Hamiltonian has purely real spectrum when its non-Hermiticity is small; at large non-Hermiticity, it changes into complex-conjugate pairs. This $\mathcal{PT}$-symmetry breaking transition occurs at an exceptional-point (EP) degeneracy where the non-orthogonal eigenmodes of the Hamiltonian also coalesce~\cite{Kato1995}. EPs occur at the ends of a branch cut of Riemann manifolds that represent complex eigenvalues, and are responsible for enhanced sensing and topological effects~\cite{Miri2019}.

These results are applicable to any system governed by a linear, first-order differential equation with a non-Hermitian generator of motion. This observation has led to the Cambrian explosion of $\mathcal{PT}$-symmetric systems, and more generally, systems with antilinear symmetries. The resulting platforms are as diverse as two waveguides~\cite{Duchesne2009,Ruter2010}, two mechanical oscillators~\cite{Bender2013}, two coupled electrical oscillators~\cite{Schindler2011,Wang2020}, two fiber loops~\cite{Regensburger2012}, two or more coupled micro-resonators~\cite{Peng2014,Hodaei2017,Chen2017}, acoustics~\cite{Zhu2014}, diffusive systems~\cite{LiY2019}, damped and driven shallow fluids~\cite{Humire2019}, and two coupled, time-delayed semiconductor lasers~\cite{Vemuri2021}. In the past two years, these classical realizations with gain and loss have been superseded by realizations in minimal quantum systems~\cite{Wu2019,Naghiloo2019,Li2019,Klauck2019} governed by post-selected, lossy Hamiltonians (or Lindbladians~\cite{Chen2021}) with similar EP degeneracies.

Fundamentally, the quantum noise in a linear amplifier~\cite{Caves1982} introduces a time-reversal asymmetry between gain and loss mechanisms~\cite{Purkayastha2020}, and makes it impossible to create $\mathcal{PT}$-symmetric systems with small  number of excitation quanta~\cite{Scheel2018}. In the semiclassical domain, customarily, the dissipation has been introduced through material impurities~\cite{Ruter2010}, coupling to a cold reservoir~\cite{Klauck2019}, or resistive elements~\cite{Schindler2011,Wang2020}; the amplification, through nonlinear gain media~\cite{Ruter2010,Regensburger2012,Peng2014,Hodaei2017}, four-wave mixing~\cite{Miri2016}, parametric driving~\cite{Roy2021}, or temporal modulation of two non-reciprocally coupled reservoirs~\cite{Hunan2020}. Balancing the gain and loss requires independent control over physically different mechanisms in spatially separated parts of the system. Therefore, engineering dynamically tunable $\mathcal{PT}$-symmetric systems is extremely challenging.


\begin{figure*}
\centering
\includegraphics[width=\textwidth]{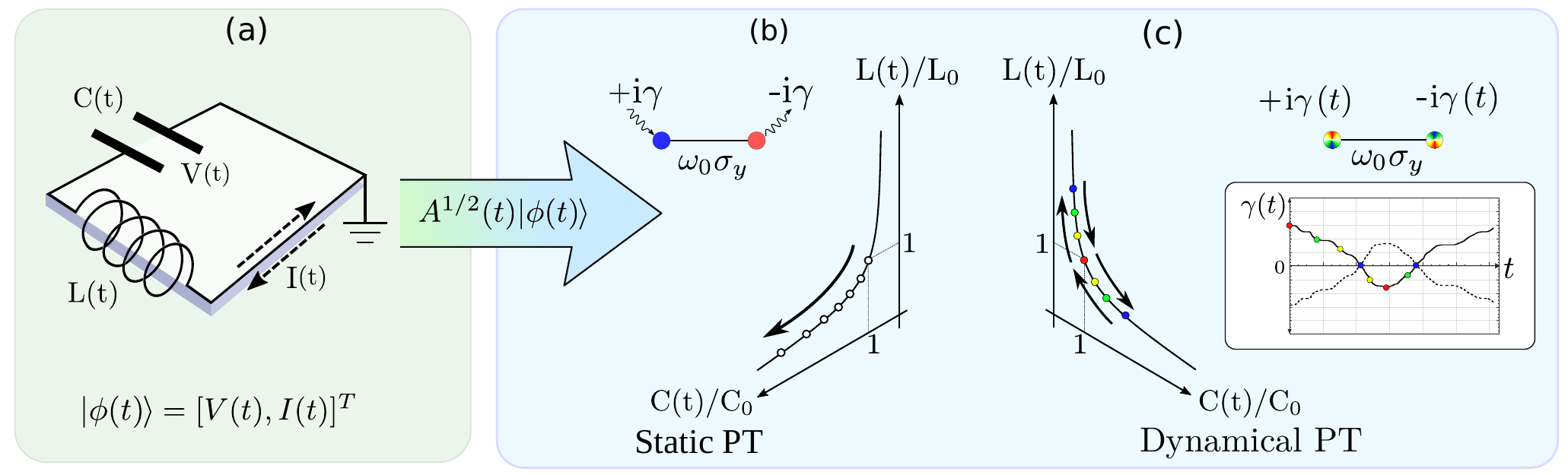}
\caption{$\mathcal{PT}$-symmetry in a single $LC$ circuit. (a) The state $|\phi(t)\rangle$ of the circuit encodes the voltage $V(t)$ across the capacitor and current $I(t)$ in the inductor, and satisfies a linear, first-order equation given by Kirchoff laws. A static change-of-basis to $|\psi(t)\rangle=A^{1/2}|\phi(t)\rangle$ leads to a Hermitian Hamiltonian $H_0=\omega_0\sigma_y$ for the new state $|\psi(t)\rangle$. Thus, an $LC$ circuit is mapped into a two-site model. (b) When $C(t), L(t)$ are exponentially varied subject to the constraint $L(t)C(t)=L_0C_0$ (hyperbola), the time-dependent change of basis generates balanced, constant gain and loss $\pm i\gamma$. This leads to a purely imaginary, $\mathcal{PT}$-symmetric Hamiltonian $H_\mathrm{PT}=\omega_0\sigma_y+i\gamma\sigma_z$. (c) When the constrained variation is not exponential,  arbitrary, but balanced gain and loss potentials $\pm i\gamma(t)$ can be generated. Here, we focus on the special case when $\gamma(t)$ is periodic.}
\label{fig:schematic}
\end{figure*}

Here, we present a radically new protocol for $\mathcal{PT}$ symmetry based on a time-dependent similarity transformation on the system of interest. The concomitant complex gauge potential, from the non-unitary change of basis, generates balanced gain and loss potentials. An added benefit of this approach is that we can implement $\mathcal{PT}$ symmetry in a minimal system with one variable i.e. a single oscillator. In contrast, all gain-loss realizations to date have required two or more coupled, separated degrees of freedom, one with the gain and the other, loss. Therefore, although the protocol is applicable to arbitrarily large systems, we will focus on the illustrative example of a single $LC$ oscillator.



\noindent{\bf $\mathcal{PT}$-symmetry in a single oscillator.} Figure~\ref{fig:schematic}a shows a schematic circuit in which the voltage $V(t)$ across the capacitor and the current $I(t)$ in the inductor satisfy Kirchhoff laws,
\begin{align}
\label{eq:v1}
I(t)+ C\frac{dV}{dt}&=0,\\
\label{eq:i1}
V(t)-L\frac{dI}{dt}&=0.
\end{align}
Equivalently, it is described by a ``state vector'' $|\phi(t)\rangle\equiv [V(t),I(t)]^T$ which satisfies the equation $i\partial_t |\phi(t)\rangle=M|\phi(t)\rangle$ where the real, non-symmetric matrix $M$ has eigenvalues $\epsilon_\pm=\pm\omega_0=\pm1/\sqrt{LC}$. Under a static, non-unitary change of basis to $|\psi(t)\rangle=A^{1/2}|\phi(t)\rangle$, the equation of motion becomes $i\partial_t |\phi(t)\rangle=H_0 |\psi(t)\rangle$ with a Hermitian Hamiltonian~\cite{LeonMontiel2018,Alan2021} $H_0=A^{1/2}MA^{-1/2}=\omega_0\sigma_y$. Here $A=\mathrm{diag}(C/2,L/2)$ is the positive-definite bilinear form that encodes the circuit energy $\mathcal{E}(t)=\langle\phi(t)|A|\phi(t)\rangle$, $\sigma_y$ is the Pauli y-matrix, and the unitary evolution under $H_0$ indicates the constancy of energy in the $LC$ circuit.

When the basis transformation depends on time, the new state vector satisfies $i\partial_t |\psi(t)\rangle=H_\mathrm{eff}(t)|\psi(t)\rangle$. The effective Hamiltonian $H_\mathrm{eff}$ is the sum of (possibly time-dependent) Hamiltonian $H_0(t)$ and a gauge potential $i\Gamma(t)$ that arises from the non-constant nature of the change of basis matrix~\cite{Nakahara2003},
\begin{align}
\label{eq:gauge}
i\Gamma(t)&=i\partial_t\ln A^{1/2}(t).
\end{align}
In fundamentally quantum systems, the matrix $A^{1/2}$ is unitary and gives rise to a Hermitian gauge term $i\Gamma=(i\Gamma)^\dagger$. In contrast, for effective models like ours, a non-unitary $A^{1/2}(t)$ can be tailored to create non-Hermitian, gain and loss potentials.

Now, let us consider constrained variations of the form $C(t)=C_0\exp[+2f(t)]$ and $L(t)=L_0\exp[-2f(t)]$ to ensure that the frequency of the oscillator remains unchanged. Such variations give rise to a traceless, anti-Hermitian gauge potential $i\Gamma=i\gamma(t)\sigma_z=i(df/dt)\sigma_z$ that represents balanced gain and loss in a solitary oscillator. The effective Hamiltonian then becomes
\begin{align}
\label{eq:heff1}
H_\mathrm{PT}(t)=\omega_0\sigma_y+i\gamma(t)\sigma_z.
\end{align}

$H_\mathrm{PT}(t)$ is invariant under combined operations of parity $\mathcal{P}=\sigma_x$ and time reversal $\mathcal{T}=*$ (complex conjugation). In contrast to quantum systems with a complex state vector, the realness of the elements of $|\psi(t)\rangle=[\sqrt{C(t)/2}V(t),\sqrt{L(t)/2}I(t)]^T$ is guaranteed by an $H_\mathrm{PT}$ with purely imaginary entries. This requirement also constrains the most general form of the non-Hermitian Hamiltonian for this system to $H_\mathrm{PT}=h_y\sigma_y+ih_z\sigma_z+ih_x\sigma_x$ with $h_k\in\mathbb{R}$.


\begin{figure*}
\centering
\includegraphics[width=\textwidth]{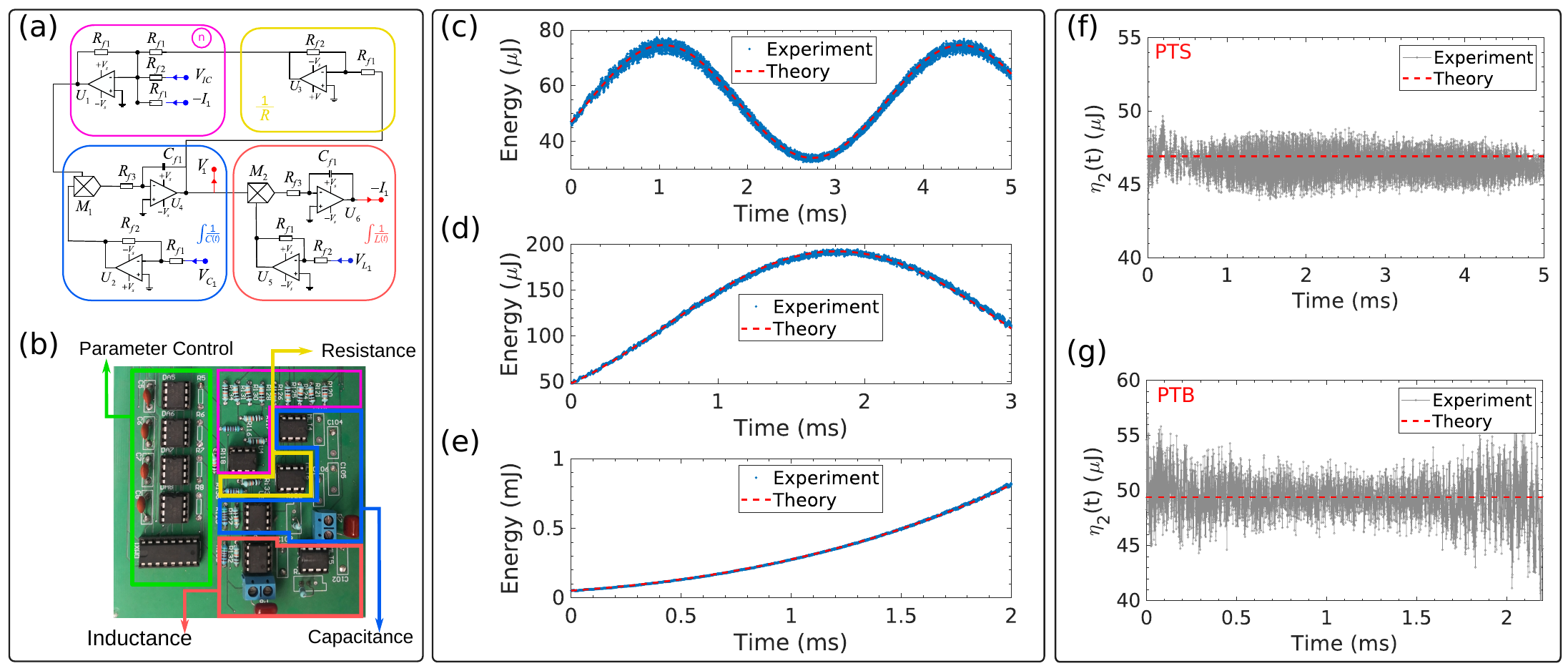}
\caption{Dynamics of a $\mathcal{PT}$-dimer with static gain and loss. (a) Schematics of synthetic $LC$ circuit comprising capacitor (red), inductor (blue), resistor (yellow), and signal adder (pink) boxes. (b) Actual circuit board with corresponding color-coded components marked.  The circuit has $C=100\,\mu\text{F}$, $L=0.01\,\text{H}$, $\omega_0=(2\pi)\times 159.15\,\text{Hz}$, and a parasitic resistance $R=10^3\,\Omega$. (c)-(d) Circuit energy $\mathcal{E}(t)$ oscillates in the $\mathcal{PT}$-symmetric phase. The gain-loss strength is (c) $\gamma=0.375\omega_0$ and (d) $\gamma=0.75\omega_0$. (e) $\mathcal{E}(t)$ grows exponentially in the $\mathcal{PT}$-broken phase, $\gamma=1.05\omega_0$ (experimental data: blue dots, theory: red dashed line). (f) At $\gamma=0.375\omega_0$ ($\mathcal{PT}$-symmetric region, PTS) although the circuit energy $\mathcal{E}(t)$ oscillates, $\eta_2(t)=\mathcal{E}(t)+\gamma V(t)I(t)/2\omega_0^2$ remains constant with time. (g) The same, constant behavior of $\eta_2(t)$ is observed at $\gamma=1.05\omega_0$ ($\mathcal{PT}$-broken region, PTB). Gray traces are experimental data; red dashed lines are theory.}
\label{fig:static}
\end{figure*}


With a suitable choice of the dimensionless function $f(t)$, Eq.(\ref{eq:heff1}) provides the protocol for arbitrary, balanced gain and loss for the energy dynamics. When $f(t)=\gamma t$ is linear in time, Eq.(\ref{eq:heff1}) gives the static $\pt$-symmetric Hamiltonian $H_\mathrm{PT}(\gamma)=\omega_0\sigma_y+i\gamma\sigma_z$. As shown in Fig.~\ref{fig:schematic}b, when $\gamma>0$, the capacitor acts as the ``gain site'' and the inductor acts as the ``loss site'' for the circuit energy.  On the other hand, surfing the hyperbola $L(t)C(t)=L_0C_0$ back and forth leads to a time-periodic $\gamma(t)$ where each ``site'' acts as a gain for fraction of the period and a loss for rest of the time (Fig.~\ref{fig:schematic}c).



\noindent{\bf Experimental results for a static $H_\mathrm{PT}$.} We experimentally demonstrate this protocol with a state-of-the-art fully reconfigurable electronic oscillator comprising functional blocks synthesized with operational amplifiers (op-amps) and passive linear components~\cite{LeonMontiel2018,Alan2020,Alan2021}. We thus electronically reproduce the dynamics described by Eqs.(\ref{eq:v1}) and (\ref{eq:i1}) in the presence of a parasitic resistance $R=10^3\,\Omega$ in parallel with the $LC$ circuit. Figure~\ref{fig:static}a shows for the circuit schematics, while the actual device is shown in Fig.~\ref{fig:static}b. $C_{0}=100\,\mu\text{F}$ is minimum capacitance and $L_{0}=0.01\,\text{H}$ is the maximum inductance that our electronic platform can efficiently simulate. Their combination gives $\omega_{0}=(2\pi)\times159\,\text{Hz}$ as the fundamental frequency of the oscillator. By increasing the capacitance at different speeds, different gain-loss strengths are realized. The eigenvalues of $H_\mathrm{PT}(\gamma)$ are $\pm\sqrt{\omega_0^2-\gamma^2}$, and they change from real to complex-conjugate pair at the exceptional point marked by $\gamma_\mathrm{EP}=\omega_0$.

The time-dependent evolution of the circuit energy $\mathcal{E}(t)=\langle\psi(t)|\psi(t)\rangle$ in the $\mathcal{PT}$-symmetric phase is shown in Fig.~\ref{fig:static} (experimental data: blue dots, theory: red dashed lines). When the gain-loss strength is doubled from $\gamma=0.375\omega_0$, Fig.~\ref{fig:static}c, to $\gamma=0.75\omega_0$, Fig.~\ref{fig:static}d, the period of oscillations increases by $\sqrt{2}$, and the amplitude of oscillations also increases. It is worth pointing out that the fast fluctuations in the experimental data are due to oscilloscope's inherent noise; as the circuit energy $\mathcal{E}(t)$ increases from tens of micro-Joules ($\mu$J) to a milli-Joule (mJ), the relative effect of the noise is suppressed. When $\gamma=1.05\omega_0$, Fig.~\ref{fig:static}e, the system goes into the $\mathcal{PT}$-broken phase, as indicated by a monotonically increasing circuit energy. The temporal range of our simulation of a static $H_\mathrm{PT}(\gamma)$ is limited the maximum value of capacitance, and not by the gain saturation of op-amps at high circuit energies.

For an ideal $\mathcal{PT}$-symmetric circuit, the energy $\mathcal{E}(t)=\langle\psi(t)|\psi(t)\rangle$ either oscillates or grows exponentially with time. And yet, for all values of $\gamma/\omega_0$, this open system has two conserved quantities given by expectation values of Hermitian, indefinite, intertwining operators~\cite{Bian2020,Ruzicka2021}. They are defined by~\cite{Mostafazadeh2002,Mostafazadeh2010}
\begin{align}
\label{eq:etaH}
\hat{\eta}_k H_\mathrm{PT}(\gamma)=H_\mathrm{PT}^\dagger(\gamma)\hat{\eta}_k,
\end{align}
and, in this case, are given by $\hat{\eta}_1=\sigma_y$ and $\hat{\eta}_2=\eta_1H_\mathrm{PT}/\omega_0=\mathbb{1}_2+(\gamma/\omega_0)\sigma_x$. The experimentally measured $\eta_2(t)\equiv\langle\psi(t)|\hat{\eta}_2|\psi(t)\rangle=\mathcal{E}(t)+\gamma V(t)I(t)/2\omega_0^2$ is shown in Fig.~\ref{fig:static}f ($\gamma=0.375\omega_0$) and Fig.~\ref{fig:static}g ($\gamma=1.05\omega_0$). Gray traces are experimental data; red dashed line is theory. $\eta_2(t)$  remains flat (modulo oscilloscope noise) in both $\mathcal{PT}$-symmetric and  $\mathcal{PT}$-broken regions. Note that the system starts out with $V(0)=0.99$ Volts and $I(0)=0$, and thus the conserved quantity $\eta_2(t)=\mathcal{E}(0)$. Since $|\psi(t)\rangle$ has real entries and $\hat{\eta}_1=\sigma_y$ has purely imaginary entries, $\eta_1(t)$ is identically equal to zero.



\begin{figure*}
\centering
\includegraphics[width=\textwidth]{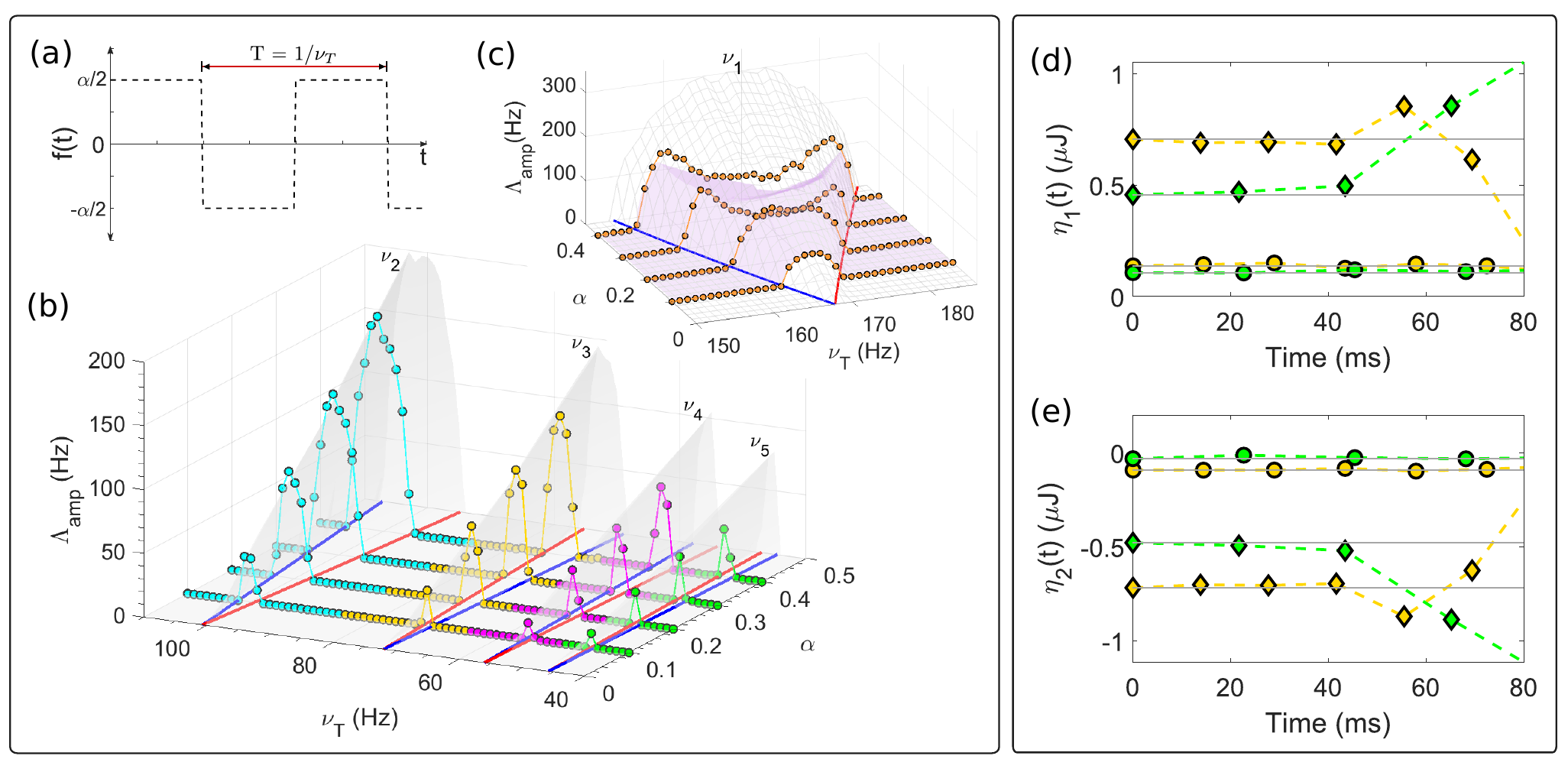}
\caption{Floquet $\mathcal{PT}$-transitions and conserved quantities. (a) The step-function $f(t)$ switches $C(t)$ and $L(t)$ by a factor of $e^{\pm\alpha}$ and gives rise to periodic, $\delta$-function gain and loss. Resultant EP lines emerging from odd resonances $\nu_k=502/(2k+1)\,\text{Hz}$ are shown by red and blue in the $\alpha-$$\nu_T$ plane. (b) $\mathcal{PT}$-broken regions  in the vicinity of $\nu_2$ (cyan), $\nu_3$ (yellow), $\nu_4$ (pink), and $\nu_5$ (green), signaled by $\Lambda_\mathrm{amp}>0$, are shown (experimental data: filled circles, theory: gray surface). (c) At $\nu_1=500/3=167\,\text{Hz}$, the gain saturation leads to a suppressed $\Lambda_\mathrm{amp}$, with the suppression largest near the resonance (experimental data: filled circles; theory with gain saturation: muve surface, theory without: gray mesh). In all cases $2\tau=50\,\text{ms}$ is used in Eq.(\ref{eq:Lambda}). (d) Constant of motion $\eta_{1F}(t_m)$ is measured near the $\nu_3$ dome (yellow) in the $\mathcal{PT}$-symmetric phase ($\nu=69\,\text{Hz}$; circles) and $\mathcal{PT}$-broken phase ($\nu=72\,\text{Hz}$; diamonds). These data are at $\alpha=0.1$. (e) Measured values of $\eta_{2F}(t_m)$ near the $\nu_5$ dome (green), both in the $\mathcal{PT}$-symmetric phase ($\nu=44\,\text{Hz}$, circles) and $\mathcal{PT}$-broken phase ($\nu=46\,\text{Hz}$, diamonds). These data are at $\alpha=0.2$. Gain saturation leads to non-constant behavior at times $t\gtrsim 2\tau$. Gray flat lines are theory in (d)-(e).}
\label{fig:floquet}
\end{figure*}



\noindent{\bf Results for a time-periodic $H_\mathrm{PT}$.} The range of dynamics generated by Eq.(\ref{eq:heff1}) is tremendously enhanced if the anti-Hermitian term $\gamma(t)$ is periodic with period $T$~\cite{Hanggi1998,Luo2013,Joglekar2014,Lee2015,Chitsazi2017,Harter2020,Akhil2021}. Then the time-evolution operator $G(t)$ at time $t=nT+\theta$ is given by $G(t)=K(\theta)G_F(T)^n$ where $K(\theta)=\mathbb{T}\exp[-i\int_0^\theta H_\mathrm{PT}(t')dt']$ captures the micromotion that occurs during a single period $0\leq\theta<T$, $\mathbb{T}$ denotes the time-ordered product, $n$ is an integer, and  $G_F(T)\equiv K(T)=\exp(-iTH_F)$ is the one-period time evolution operator, that, in turn, defines the Floquet Hamiltonian $H_F$. The complex eigenvalues $\lambda_\pm$ of $G_F(T)$ determine whether the system is in the $\mathcal{PT}$ symmetric region ($|\lambda_+|=|\lambda_-|$) or broken region ($|\lambda_+|\neq|\lambda_-|$).

Congruent with the experimental setup, we use the function $f(t)=\alpha\Pi(t)=f(t+T)$ where $\Pi(t)=\mathrm{sgn}(t)/2$ for $|t|\leq T/2$ is the unit-step square wave (Fig.~\ref{fig:floquet}a), and $\alpha$ quantifies the extent of constrained variation, i.e. $ e^{-\alpha}\leq C(t)/C_0, L(t)/L_0\leq e^{\alpha}$. By taking into account the $\delta$-function  generated by $\partial_t\Pi(t)$, it is straightforward to evaluate the purely real, one-period operator
\begin{align}
\label{eq:Gf2}
G_F(T)&= e^{+\alpha\sigma_z}e^{-i\omega_0\sigma_yT/2}e^{-\alpha\sigma_z}e^{-i\omega_0\sigma_yT/2}.
\end{align}
Its eigenvalues are $\lambda_\pm=C^2-S^2\cosh(2\alpha)\pm iS\sqrt{D}$ where the discriminant is given by $D= C^2[1+\cosh(2\alpha)]^2-\sinh^2(2\alpha)$, and $C=\cos(\omega_0T/2)$, $S=\sin(\omega_0T/2)$. The boundary between the $\mathcal{PT}$-symmetric and $\mathcal{PT}$-broken regions in the $\alpha-$$\nu_T$ plane ($\nu_T=1/T$) is marked by a vanishing discriminant $D=0$ or, equivalently,
\begin{equation}
\label{eq:Gf4}
\cos(\omega_0T/2)=\pm\tanh(\alpha_\mathrm{EP}).
\end{equation}

At $\alpha=0$, the eigenvalues $\lambda_\pm=e^{\pm i\omega_0T}$ of the matrix $G_F$ become degenerate at odd resonances $2\pi\nu_n=2\omega_0/(2n+1)$. These are diabolic-point (DP) degeneracies. At small $\alpha$ symmetrical EP lines, emerging from the DP, satisfy the equation $\delta\nu_n(\alpha_\mathrm{EP})=\pm A_n\alpha_\mathrm{EP}$ where $\delta\nu_n=\nu_T-\nu_n$ is the distance from DP and $A_n=2\omega_0/[(2n+1)\pi]^2$. Thus, the $\mathcal{PT}$-broken region at arbitrarily small $\alpha$, bounded by the two EP lines, becomes narrower with increasing $n$~\cite{Harter2020,Akhil2021}. Figures~\ref{fig:floquet}b,c show these lines in the $\alpha-$$\nu_T$ plane.

For this set of experiments, using $C_{0}=400\,\mu\text{F}$ and $L_{0}=1\,\text{mH}$ fixes the oscillator frequency at $\omega_0/(2\pi)=251\,\text{Hz}$. We use an experimentally friendly parameter~\cite{LeonMontiel2018,Cochran2021}
\begin{align}
\label{eq:Lambda}
\Lambda_\mathrm{amp}(\alpha,\nu_T) =\lim_{2\tau\gg T}\frac{1}{\tau}\log\left[\frac{\max\mathcal{E}(0\leq t\leq2\tau)}{\max\mathcal{E}(0\leq t\leq\tau)}\right],
\end{align}
obtained from the circuit energy to characterize the strength of the $\mathcal{PT}$-broken phase. Since $\mathcal{E}(t)$ oscillates in the $\mathcal{PT}$-symmetric phase, $\Lambda_\mathrm{amp}=0$, whereas its exponential growth in the $\mathcal{PT}$-broken region gives $\Lambda_\mathrm{amp}>0$. Figure~\ref{fig:floquet}b shows the emergent triangular $\pt$-broken regions at $\nu_2=100\,\text{Hz}$ (cyan circles), $\nu_3=71\,\text{Hz}$ (yellow circles), $\nu_4=55\,\text{Hz}$ (pink circles), and $\nu_5=45\,\text{Hz}$ (green circles). The gray surface is theory. In the $\mathcal{PT}$-broken regions, at high circuit energies, $\mathcal{E}(t)$ does not grow exponentially due to op-amp saturation. This leads to a suppression of the effective amplification rate $\Lambda_\mathrm{amp}$. This suppression is maximum in the deepest $\mathcal{PT}$-symmetry broken region and leads to $\Lambda_\mathrm{amp}\rightarrow 0$ as the circuit energy saturates at short times. It is clearly seen in Fig.~\ref{fig:floquet}c at $\nu_1=167\,\text{Hz}$ (orange circles) at large $\alpha$, but is absent at smaller values of $\alpha$. The mauve surface is a theory prediction with gain saturation $+i\gamma(V)$, whereas the gray mesh is one without gain saturation. This suppression is almost complete at the primary resonance which occurs at $\nu_0=2\omega_0/(2\pi)=502\,\text{Hz}$ (not shown) limiting the use of Eq.(\ref{eq:Lambda}) to distinguish between $\mathcal{PT}$-symmetric and broken phases.

In the Floquet case, the conserved quantities are the expectation values of Hermitian, indefinite operators $\hat{\eta}_F$ that satisfy the intertwining relation $G^\dagger_F(T)\hat{\eta}_F G_F(T)=\hat{\eta}_F$, i.e. $\eta_F(t_m)\equiv\langle\psi(mT)|\hat{\eta}_F|\psi(mT)\rangle$ is independent of $m$~\cite{Ruzicka2021}. We choose them as $\hat{\eta}_{1F}=\cosh(\alpha)\mathbb{1}_2-\sinh(\alpha)\sigma_z+\sinh(\alpha)\tan(\omega_0T/2)\sigma_x$ and $\hat{\eta}_{2F}=\hat{\eta}_{1F}G_F(T)$. Figure~\ref{fig:floquet}d shows $\eta_{1F}$ obtained from the experimentally measured state-vector $|\psi(t)\rangle$ at times $t=t_m$. Since $\nu\sim70\,\text{Hz}-40\,\text{Hz}$, there are $\sim5$ stroboscopic data points available. These are representative results in the $\mathcal{PT}$-symmetric (circles) and $\mathcal{PT}$-broken (diamonds) phases in the vicinity of $\nu_3$ and $\nu_5$ domes with color-coded symbols (Fig.~\ref{fig:floquet}b). Experimentally measured $\eta_{2F}(t_m)$ results are shown in Fig.~\ref{fig:floquet}e. The deviation from flat gray lines (theory) at long times $t\gtrsim 2\tau$ is due to gain saturation in the $\mathcal{PT}$-broken phase (diamonds). Since $G_F(T)=-\mathbb{1}_2$ at the DP degeneracies $\nu_k=502/(2k+1)\,\text{Hz}$, we see that $\eta_{2F}(t_m)=-\eta_{1F}(t_m)$ holds, surprisingly, irrespective of the gain saturation. Similar results are valid, of course, across the entire $\alpha-$$\nu_T$ plane. This remarkable ability to map out the entire Floquet $\mathcal{PT}$-phase diagram across five domes showcases the tremendous versatility of the synthetic electronic platform, and the distinct advantage of complex gauge-field induced gain and loss mechanism over traditional approaches~\cite{Ruter2010,Schindler2011,Wang2020,Chitsazi2017,Regensburger2012,Peng2014,Hodaei2017,Klauck2019,Roy2021,Hunan2020,Miri2016}. This unparalleled versatility has also enabled the first demonstration of conserved quantities in the Floquet dynamics of a $\mathcal{PT}$-symmetric system.




\begin{figure*}
\begin{center}
\includegraphics[width=\textwidth]{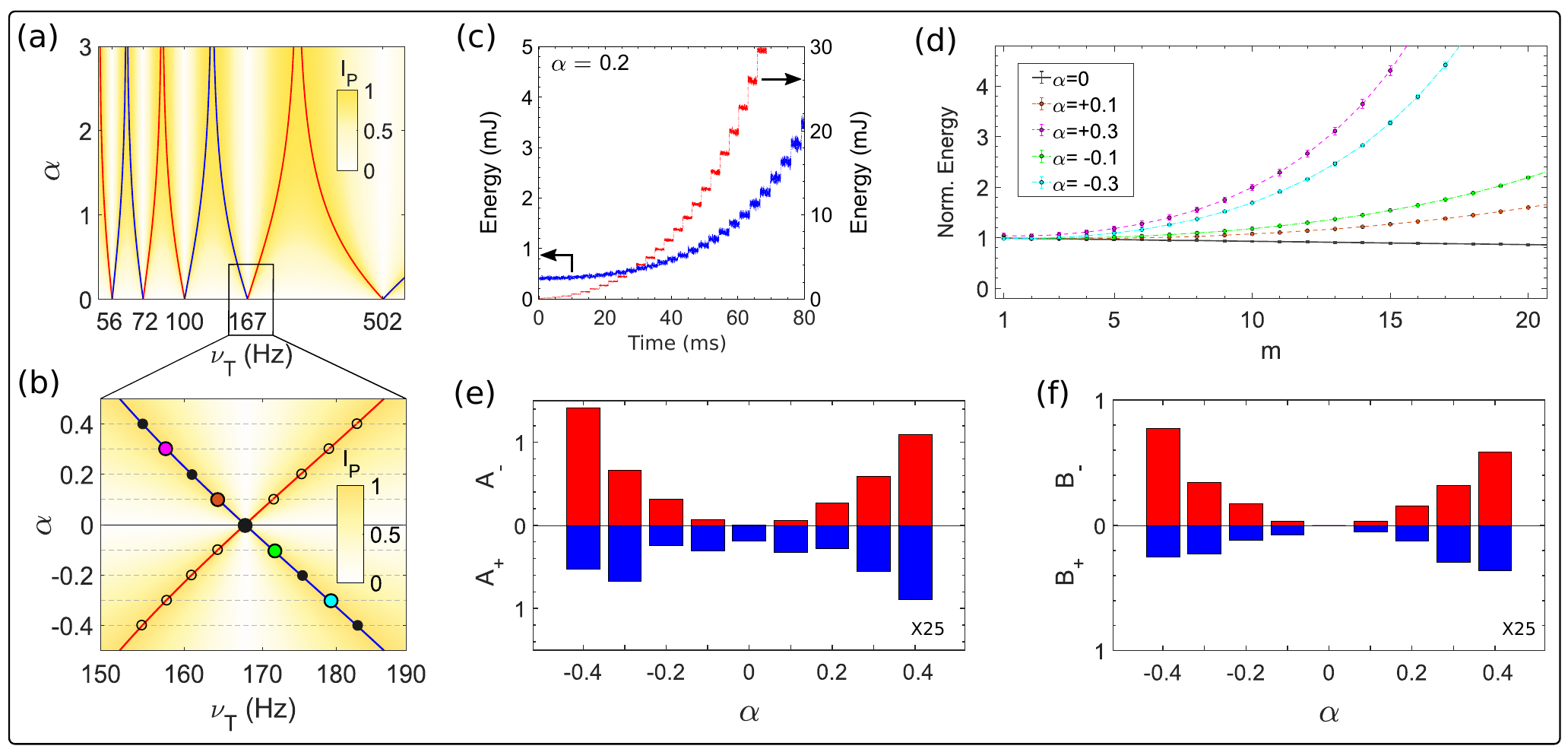}
\end{center}
\caption{Giant dynamical asymmetry along the EP contours.  (a) The inner-product heat map $I_P(\alpha,\nu_T)$ shows EP lines emerging from odd resonances $\nu_k=502/(2k+1)\,\text{Hz}$, consistent with  Eq.(\ref{eq:Gf4}). (b) Zoomed-in view near $\nu_1=167\,\text{Hz}$ shows the $\alpha_\mathrm{EP}$ values sampled for the circuit energy dynamics (filled and open circles). (c) Experimentally measured $\mathcal{E}(t)$ for the two EPs at $\alpha=0.2$ show a giant asymmetry. The circuit energy is constant except at times $pT/2$ ($p\geq 1$) when the $\delta$-function gain-loss potentials are active. (d) Color-coded stroboscopic, normalized energy traces along the blue contour show quadratic behavior consistent with a second-order EP. The (constant) energy along the flat steps in (c) is averaged to obtain error bars  on $\mathcal{E}(t_m)/\mathcal{E}(0)$. By fitting the data to Eq.(\ref{eq:epline}), $A_\pm(\alpha), B_\pm(\alpha)$ are obtained for the nine $\alpha$ values sampled  along blue and red contours. (e) $A_\pm(\alpha)$ show approximate $\alpha\leftrightarrow-\alpha$ symmetry, vanish at the DP as expected,
and show that $\mathcal{E}(t_m)$ growth along the red contour is $\sim25$-fold larger than along the blue contour. (f) Similar results, including a giant 25-fold asymmetry, are obtained for $B_\pm(\alpha)$.}
\label{fig:epline}
\end{figure*}


\noindent{\bf Walking the EP contours.} As a last demonstration of the synthetic oscillator platform, we investigate the temporal dynamics at numerous points along the EP contours. This has been extremely challenging in gain-loss systems due to the requisite fine-tuning of multiple mechanisms. In loss-only $\mathcal {PT}$-systems, it is a challenge because decay rate is maximum at the EP.

A complementary way to show the $\mathcal{PT}$-phase diagram in the $\alpha-$$\nu_T$ plane is the Dirac inner-product of the right eigenvectors of $G_F(T)$. By expressing the real, Floquet evolution matrix as $G_F(T)=G_0\mathbb{1}_2+G_x\sigma_x+G_z\sigma_z+iG_y\sigma_y$ where $G_k\in\mathbb{R}$, it is straightforward to obtain the inner-product as $I_P(\alpha,\nu_T)=\min(r,1/r)$ where
\begin{align}
r&=\frac{G_y^2}{\sqrt{G_x^2+G_z^2}}=\frac{\tanh(\alpha)}{\cos(\omega_0/2\nu_T)}.
\end{align}

Figure~\ref{fig:epline}a shows the heat-map overlaid with EP contours, $I_P=1$, where blue is for the plus sign and red corresponds to the minus sign in Eq.(\ref{eq:Gf4}). Figure~\ref{fig:epline}b shows the region near $\nu_1=167\,\text{Hz}$ extended to negative values of $\alpha$. We ``park the system'' at 9 points with equidistant $\alpha_\mathrm{EP}$ values along the blue (filled circles) and red (open circles) contours each, and obtain the circuit energy evolution $\mathcal{E}(t)$. Since the system has second-order EP contours, the stroboscopic circuit energy $\mathcal{E}(t_m)=\langle\psi(0)|G_F^\dagger(mT)G_F(mT)|\psi(0)\rangle$ grows quadratically with time $t_m$.  Figure~\ref{fig:epline}c shows the experimentally measured circuit energy $\mathcal{E}(t)$ at $\alpha=0.2$ on the blue contour  (blue trace) and the red contour (red trace) over $m\sim 25$ periods. It is constant except at integer and half-integer periods when the $\delta$-function gain-loss potential is active. Surprisingly, $\mathcal{E}(t)$ also shows an order-of-magnitude asymmetry for the two contours that emerge symmetrically from the DP at $\nu_1$; this asymmetry persists at all $\alpha_\mathrm{EP}$.

We quantify the growth of stroboscopic, normalized circuit energy $\mathcal{E}(t_m)$ with two dimensionless coefficients,
\begin{align}
\label{eq:epline}
\frac{\mathcal{E}(t_m)}{\mathcal{E}(0)}=1-A_\pm(\alpha_\mathrm{EP}) m+ B_\pm(\alpha_\mathrm{EP})m^2,
\end{align}
that depend only on $\alpha_\mathrm{EP}$ since it uniquely determines the corresponding $\nu_T$ via Eq.(\ref{eq:Gf4}). This approach allows us to investigate their dependence on the proximity to the DP degeneracy at $\alpha=0$, and the dynamical asymmetry between temporal evolution along the blue (plus) and red (minus) EP contours. At the second-order EP,  $G_F(t_m)=e^{-ih_0T}(\mathbb{1}_2-imTH_F)$ where $h_0T=\pi$ and $H_F=[\mp2\sinh(\alpha_\mathrm{EP})\sigma_y+2i\tanh(\alpha_\text{EP})\sigma_z\pm 2i\sinh(\alpha_\mathrm{EP})\tanh(\alpha_\mathrm{EP})\sigma_x]/T$ is the Floquet Hamiltonian. We obtain $A=iT\langle\psi(0)|H_F^{\dagger}-H_F|\psi(0)\rangle/\mathcal{E}(0)$ and  $B=T^2\langle\psi(0)|H_F^{\dagger} H_F|\psi(0)\rangle/\mathcal{E}(0)\geq 0$  for the coefficients in Eq.(\ref{eq:epline}). Figure~\ref{fig:epline}d shows the stroboscopic, normalized circuit energy $\mathcal{E}(t_m)/\mathcal{E}(0)$ as a function of $m$ along the blue contour for $|\alpha|=\{0,0.1,0.3\}$. The error bars on $\mathcal{E}(t_m)$ are obtained by averaging its value over the constant region. As the DP at $\alpha=0$ is approached from either side,  the coefficients $A_{+}(\alpha)$ and $B_{+}(\alpha)$ are monotonically suppressed to zero. The slight negative slope of $\mathcal{E}(t_m)$ at $\alpha=0$ (no gain or loss) is due to the parasitic resistance in the circuit.

We extract the coefficients $A_{-}(\alpha)$ (Fig.~\ref{fig:epline}e) and $B_{-}(\alpha)$ (Fig.~\ref{fig:epline}f) from the experimental data along the red contour. Similar results, with dramatically smaller values of $A_{+}$ and $B_{+}$, are obtained for a walk along the blue contour. They clearly demonstrate the order-of-magnitude dynamical asymmetry that arises when the same initial state $|\psi(0)\rangle$, with a fully charged capacitor, is evolved along the two symmetrical EP lines that emerge from $\nu_1=167\,\text{Hz}$.


\noindent{\bf Discussion.} Most  of the transformative ideas in non-Hermitian physics---Riemann surfaces, bi-orthogonal basis, exceptional points, to name a few---have been well-known in mathematics. Yet, their reinterpreation in the context of open systems has lent novel insights that underpin the advances such as enhanced sensing~\cite{Chen2017,Hodaei2017}, chiral mode switch~\cite{Doppler2016,Xu2016}, or topological braiding~\cite{KWang2021}.

Similarly, starting with the Kanai model~\cite{Bateman1931,Kanai1948}, a quantum harmonic oscillator with exponentially varying mass~\cite{Britten1950} has been extensively studied~\cite{Feshbach1977,Green1979a,Green1979b,Dekker1981}, with a focus on the fate of the uncertainty relation~\cite{Cheng1988} and non-unitary canonical transformations~\cite{Gomez2007}.

We have, instead, contextualized time-dependent non-unitary transformations into a novel protocol to implement balanced gain and loss in a single oscillator. Simple harmonic oscillator based models are all-pervasive in nature, and our protocol provides a recipe for their non-Hermitian generalization. For example, in a metamaterial, the non-unitary change of basis is given by the permittivity $\epsilon$ and the permeability $\mu$. A constrained variation of the two, with a constant product (and therefore a constant index of refraction), can lead to a new class of $\mathcal{PT}$-symmetric metamaterials~\cite{Sotiris2019} without material gain or loss. It is also easy to generalize this protocol to a network of oscillators, where the gain and loss ``sites'' are localized in different nodes or are distributed throughout the network. For example,  with reconfigurable synthetic $LC$ circuits, this method can lead to non-passive, $\mathcal{PT}$-symmetric extensions of topoelectrical circuits~\cite{Lee2018}.

\vspace{5mm}
\noindent{\it Acknowledgments.} This work was supported by DGAPA-UNAM under the project UNAM-PAPIIT IN102920, and by CONACYT under the projects No. CB-2016-01/284372 and No. A1-S-8317.


\bibliography{biblio}

\begin{thebibliography}{61}%
\makeatletter
\providecommand \@ifxundefined [1]{%
 \@ifx{#1\undefined}
}%
\providecommand \@ifnum [1]{%
 \ifnum #1\expandafter \@firstoftwo
 \else \expandafter \@secondoftwo
 \fi
}%
\providecommand \@ifx [1]{%
 \ifx #1\expandafter \@firstoftwo
 \else \expandafter \@secondoftwo
 \fi
}%
\providecommand \natexlab [1]{#1}%
\providecommand \enquote  [1]{``#1''}%
\providecommand \bibnamefont  [1]{#1}%
\providecommand \bibfnamefont [1]{#1}%
\providecommand \citenamefont [1]{#1}%
\providecommand \href@noop [0]{\@secondoftwo}%
\providecommand \href [0]{\begingroup \@sanitize@url \@href}%
\providecommand \@href[1]{\@@startlink{#1}\@@href}%
\providecommand \@@href[1]{\endgroup#1\@@endlink}%
\providecommand \@sanitize@url [0]{\catcode `\\12\catcode `\$12\catcode
  `\&12\catcode `\#12\catcode `\^12\catcode `\_12\catcode `\%12\relax}%
\providecommand \@@startlink[1]{}%
\providecommand \@@endlink[0]{}%
\providecommand \url  [0]{\begingroup\@sanitize@url \@url }%
\providecommand \@url [1]{\endgroup\@href {#1}{\urlprefix }}%
\providecommand \urlprefix  [0]{URL }%
\providecommand \Eprint [0]{\href }%
\providecommand \doibase [0]{https://doi.org/}%
\providecommand \selectlanguage [0]{\@gobble}%
\providecommand \bibinfo  [0]{\@secondoftwo}%
\providecommand \bibfield  [0]{\@secondoftwo}%
\providecommand \translation [1]{[#1]}%
\providecommand \BibitemOpen [0]{}%
\providecommand \bibitemStop [0]{}%
\providecommand \bibitemNoStop [0]{.\EOS\space}%
\providecommand \EOS [0]{\spacefactor3000\relax}%
\providecommand \BibitemShut  [1]{\csname bibitem#1\endcsname}%
\let\auto@bib@innerbib\@empty
\bibitem [{\citenamefont {Bender}\ and\ \citenamefont
  {Boettcher}(1998)}]{Bender1998}%
  \BibitemOpen
  \bibfield  {author} {\bibinfo {author} {\bibfnamefont {C.~M.}\ \bibnamefont
  {Bender}}\ and\ \bibinfo {author} {\bibfnamefont {S.}~\bibnamefont
  {Boettcher}},\ }\bibfield  {title} {\bibinfo {title} {Real spectra in
  non-hermitian hamiltonians having $\mathcal{PT}$ symmetry},\ }\href
  {https://doi.org/10.1103/PhysRevLett.80.5243} {\bibfield  {journal} {\bibinfo
   {journal} {Phys. Rev. Lett.}\ }\textbf {\bibinfo {volume} {80}},\ \bibinfo
  {pages} {5243} (\bibinfo {year} {1998})}\BibitemShut {NoStop}%
\bibitem [{\citenamefont {Bender}\ \emph {et~al.}(2002)\citenamefont {Bender},
  \citenamefont {Brody},\ and\ \citenamefont {Jones}}]{Bender2001}%
  \BibitemOpen
  \bibfield  {author} {\bibinfo {author} {\bibfnamefont {C.~M.}\ \bibnamefont
  {Bender}}, \bibinfo {author} {\bibfnamefont {D.~C.}\ \bibnamefont {Brody}},\
  and\ \bibinfo {author} {\bibfnamefont {H.~F.}\ \bibnamefont {Jones}},\
  }\bibfield  {title} {\bibinfo {title} {Complex extension of quantum
  mechanics},\ }\href {https://doi.org/10.1103/PhysRevLett.89.270401}
  {\bibfield  {journal} {\bibinfo  {journal} {Phys. Rev. Lett.}\ }\textbf
  {\bibinfo {volume} {89}},\ \bibinfo {pages} {270401} (\bibinfo {year}
  {2002})}\BibitemShut {NoStop}%
\bibitem [{\citenamefont {Duchesne}\ \emph {et~al.}(2009)\citenamefont
  {Duchesne}, \citenamefont {Aimez}, \citenamefont {Morandotti}, \citenamefont
  {Christodoulides}, \citenamefont {Salamo}, \citenamefont {Volatier-Ravat},
  \citenamefont {Siviloglou},\ and\ \citenamefont {Guo}}]{Duchesne2009}%
  \BibitemOpen
  \bibfield  {author} {\bibinfo {author} {\bibfnamefont {D.}~\bibnamefont
  {Duchesne}}, \bibinfo {author} {\bibfnamefont {V.}~\bibnamefont {Aimez}},
  \bibinfo {author} {\bibfnamefont {R.}~\bibnamefont {Morandotti}}, \bibinfo
  {author} {\bibfnamefont {D.~N.}\ \bibnamefont {Christodoulides}}, \bibinfo
  {author} {\bibfnamefont {G.~J.}\ \bibnamefont {Salamo}}, \bibinfo {author}
  {\bibfnamefont {M.}~\bibnamefont {Volatier-Ravat}}, \bibinfo {author}
  {\bibfnamefont {G.~A.}\ \bibnamefont {Siviloglou}},\ and\ \bibinfo {author}
  {\bibfnamefont {A.}~\bibnamefont {Guo}},\ }\bibfield  {title} {\bibinfo
  {title} {{Observation of PT -Symmetry Breaking in Complex Optical
  Potentials}},\ }\href {https://doi.org/10.1103/physrevlett.103.093902}
  {\bibfield  {journal} {\bibinfo  {journal} {Physical Review Letters}\
  }\textbf {\bibinfo {volume} {103}},\ \bibinfo {pages} {1} (\bibinfo {year}
  {2009})}\BibitemShut {NoStop}%
\bibitem [{\citenamefont {R{\"{u}}ter}\ \emph {et~al.}(2010)\citenamefont
  {R{\"{u}}ter}, \citenamefont {Makris}, \citenamefont {El-Ganainy},
  \citenamefont {Christodoulides}, \citenamefont {Segev},\ and\ \citenamefont
  {Kip}}]{Ruter2010}%
  \BibitemOpen
  \bibfield  {author} {\bibinfo {author} {\bibfnamefont {C.~E.}\ \bibnamefont
  {R{\"{u}}ter}}, \bibinfo {author} {\bibfnamefont {K.~G.}\ \bibnamefont
  {Makris}}, \bibinfo {author} {\bibfnamefont {R.}~\bibnamefont {El-Ganainy}},
  \bibinfo {author} {\bibfnamefont {D.~N.}\ \bibnamefont {Christodoulides}},
  \bibinfo {author} {\bibfnamefont {M.}~\bibnamefont {Segev}},\ and\ \bibinfo
  {author} {\bibfnamefont {D.}~\bibnamefont {Kip}},\ }\bibfield  {title}
  {\bibinfo {title} {{Observation of parity-time symmetry in optics}},\ }\href
  {https://doi.org/10.1038/nphys1515} {\bibfield  {journal} {\bibinfo
  {journal} {Nature Physics}\ }\textbf {\bibinfo {volume} {6}},\ \bibinfo
  {pages} {192} (\bibinfo {year} {2010})},\ \Eprint
  {https://arxiv.org/abs/1003.4968} {1003.4968} \BibitemShut {NoStop}%
\bibitem [{\citenamefont {Feng}\ \emph {et~al.}(2017)\citenamefont {Feng},
  \citenamefont {El-Ganainy},\ and\ \citenamefont {Ge}}]{Feng2017}%
  \BibitemOpen
  \bibfield  {author} {\bibinfo {author} {\bibfnamefont {L.}~\bibnamefont
  {Feng}}, \bibinfo {author} {\bibfnamefont {R.}~\bibnamefont {El-Ganainy}},\
  and\ \bibinfo {author} {\bibfnamefont {L.}~\bibnamefont {Ge}},\ }\bibfield
  {title} {\bibinfo {title} {Non-hermitian photonics based on
  parity{\textendash}time symmetry},\ }\href
  {https://doi.org/10.1038/s41566-017-0031-1} {\bibfield  {journal} {\bibinfo
  {journal} {Nature Photonics}\ }\textbf {\bibinfo {volume} {11}},\ \bibinfo
  {pages} {752} (\bibinfo {year} {2017})}\BibitemShut {NoStop}%
\bibitem [{\citenamefont {El-Ganainy}\ \emph {et~al.}(2018)\citenamefont
  {El-Ganainy}, \citenamefont {Makris}, \citenamefont {Khajavikhan},
  \citenamefont {Musslimani}, \citenamefont {Rotter},\ and\ \citenamefont
  {Christodoulides}}]{ElGanainy2018}%
  \BibitemOpen
  \bibfield  {author} {\bibinfo {author} {\bibfnamefont {R.}~\bibnamefont
  {El-Ganainy}}, \bibinfo {author} {\bibfnamefont {K.~G.}\ \bibnamefont
  {Makris}}, \bibinfo {author} {\bibfnamefont {M.}~\bibnamefont {Khajavikhan}},
  \bibinfo {author} {\bibfnamefont {Z.~H.}\ \bibnamefont {Musslimani}},
  \bibinfo {author} {\bibfnamefont {S.}~\bibnamefont {Rotter}},\ and\ \bibinfo
  {author} {\bibfnamefont {D.~N.}\ \bibnamefont {Christodoulides}},\ }\bibfield
   {title} {\bibinfo {title} {Non-hermitian physics and {PT} symmetry},\ }\href
  {https://doi.org/10.1038/nphys4323} {\bibfield  {journal} {\bibinfo
  {journal} {Nature Physics}\ }\textbf {\bibinfo {volume} {14}},\ \bibinfo
  {pages} {11} (\bibinfo {year} {2018})}\BibitemShut {NoStop}%
\bibitem [{\citenamefont {Gupta}\ \emph {et~al.}(2019)\citenamefont {Gupta},
  \citenamefont {Zou}, \citenamefont {Zhu}, \citenamefont {Lu}, \citenamefont
  {Zhang}, \citenamefont {Liu},\ and\ \citenamefont {Chen}}]{Gupta2019}%
  \BibitemOpen
  \bibfield  {author} {\bibinfo {author} {\bibfnamefont {S.~K.}\ \bibnamefont
  {Gupta}}, \bibinfo {author} {\bibfnamefont {Y.}~\bibnamefont {Zou}}, \bibinfo
  {author} {\bibfnamefont {X.-Y.}\ \bibnamefont {Zhu}}, \bibinfo {author}
  {\bibfnamefont {M.-H.}\ \bibnamefont {Lu}}, \bibinfo {author} {\bibfnamefont
  {L.-J.}\ \bibnamefont {Zhang}}, \bibinfo {author} {\bibfnamefont {X.-P.}\
  \bibnamefont {Liu}},\ and\ \bibinfo {author} {\bibfnamefont {Y.-F.}\
  \bibnamefont {Chen}},\ }\bibfield  {title} {\bibinfo {title} {Parity-time
  symmetry in non-hermitian complex optical media},\ }\href
  {https://doi.org/10.1002/adma.201903639} {\bibfield  {journal} {\bibinfo
  {journal} {Advanced Materials}\ ,\ \bibinfo {pages} {1903639}} (\bibinfo
  {year} {2019})}\BibitemShut {NoStop}%
\bibitem [{\citenamefont {Kato}(1995)}]{Kato1995}%
  \BibitemOpen
  \bibfield  {author} {\bibinfo {author} {\bibfnamefont {T.}~\bibnamefont
  {Kato}},\ }\href {https://doi.org/10.1007/978-3-642-66282-9} {\emph {\bibinfo
  {title} {Perturbation Theory for Linear Operators}}}\ (\bibinfo  {publisher}
  {Springer Berlin Heidelberg},\ \bibinfo {year} {1995})\BibitemShut {NoStop}%
\bibitem [{\citenamefont {Miri}\ and\ \citenamefont
  {Al{\`{u}}}(2019)}]{Miri2019}%
  \BibitemOpen
  \bibfield  {author} {\bibinfo {author} {\bibfnamefont {M.-A.}\ \bibnamefont
  {Miri}}\ and\ \bibinfo {author} {\bibfnamefont {A.}~\bibnamefont
  {Al{\`{u}}}},\ }\bibfield  {title} {\bibinfo {title} {Exceptional points in
  optics and photonics},\ }\href {https://doi.org/10.1126/science.aar7709}
  {\bibfield  {journal} {\bibinfo  {journal} {Science}\ }\textbf {\bibinfo
  {volume} {363}},\ \bibinfo {pages} {eaar7709} (\bibinfo {year}
  {2019})}\BibitemShut {NoStop}%
\bibitem [{\citenamefont {Bender}\ \emph {et~al.}(2013)\citenamefont {Bender},
  \citenamefont {Berntson}, \citenamefont {Parker},\ and\ \citenamefont
  {Samuel}}]{Bender2013}%
  \BibitemOpen
  \bibfield  {author} {\bibinfo {author} {\bibfnamefont {C.~M.}\ \bibnamefont
  {Bender}}, \bibinfo {author} {\bibfnamefont {B.~K.}\ \bibnamefont
  {Berntson}}, \bibinfo {author} {\bibfnamefont {D.}~\bibnamefont {Parker}},\
  and\ \bibinfo {author} {\bibfnamefont {E.}~\bibnamefont {Samuel}},\
  }\bibfield  {title} {\bibinfo {title} {Observation of {PT} phase transition
  in a simple mechanical system},\ }\href {https://doi.org/10.1119/1.4789549}
  {\bibfield  {journal} {\bibinfo  {journal} {American Journal of Physics}\
  }\textbf {\bibinfo {volume} {81}},\ \bibinfo {pages} {173} (\bibinfo {year}
  {2013})}\BibitemShut {NoStop}%
\bibitem [{\citenamefont {Schindler}\ \emph {et~al.}(2011)\citenamefont
  {Schindler}, \citenamefont {Li}, \citenamefont {Zheng}, \citenamefont
  {Ellis},\ and\ \citenamefont {Kottos}}]{Schindler2011}%
  \BibitemOpen
  \bibfield  {author} {\bibinfo {author} {\bibfnamefont {J.}~\bibnamefont
  {Schindler}}, \bibinfo {author} {\bibfnamefont {A.}~\bibnamefont {Li}},
  \bibinfo {author} {\bibfnamefont {M.~C.}\ \bibnamefont {Zheng}}, \bibinfo
  {author} {\bibfnamefont {F.~M.}\ \bibnamefont {Ellis}},\ and\ \bibinfo
  {author} {\bibfnamefont {T.}~\bibnamefont {Kottos}},\ }\bibfield  {title}
  {\bibinfo {title} {{Experimental study of active LRC circuits with PT
  symmetries}},\ }\href {https://doi.org/10.1103/PhysRevA.84.040101} {\bibfield
   {journal} {\bibinfo  {journal} {Physical Review A - Atomic, Molecular, and
  Optical Physics}\ }\textbf {\bibinfo {volume} {84}},\ \bibinfo {pages} {1}
  (\bibinfo {year} {2011})}\BibitemShut {NoStop}%
\bibitem [{\citenamefont {Wang}\ \emph {et~al.}(2020)\citenamefont {Wang},
  \citenamefont {Fang}, \citenamefont {Xie}, \citenamefont {Dong},
  \citenamefont {Joglekar}, \citenamefont {Wang}, \citenamefont {Li},\ and\
  \citenamefont {Luo}}]{Wang2020}%
  \BibitemOpen
  \bibfield  {author} {\bibinfo {author} {\bibfnamefont {T.}~\bibnamefont
  {Wang}}, \bibinfo {author} {\bibfnamefont {J.}~\bibnamefont {Fang}}, \bibinfo
  {author} {\bibfnamefont {Z.}~\bibnamefont {Xie}}, \bibinfo {author}
  {\bibfnamefont {N.}~\bibnamefont {Dong}}, \bibinfo {author} {\bibfnamefont
  {Y.~N.}\ \bibnamefont {Joglekar}}, \bibinfo {author} {\bibfnamefont
  {Z.}~\bibnamefont {Wang}}, \bibinfo {author} {\bibfnamefont {J.}~\bibnamefont
  {Li}},\ and\ \bibinfo {author} {\bibfnamefont {L.}~\bibnamefont {Luo}},\
  }\bibfield  {title} {\bibinfo {title} {Observation of two pt transitions in
  an electric circuit with balanced gain and loss},\ }\bibfield  {journal}
  {\bibinfo  {journal} {The European Physical Journal D}\ }\textbf {\bibinfo
  {volume} {74}},\ \href {https://doi.org/10.1140/epjd/e2020-10131-7}
  {10.1140/epjd/e2020-10131-7} (\bibinfo {year} {2020})\BibitemShut {NoStop}%
\bibitem [{\citenamefont {Regensburger}\ \emph {et~al.}(2012)\citenamefont
  {Regensburger}, \citenamefont {Bersch}, \citenamefont {Miri}, \citenamefont
  {Onishchukov}, \citenamefont {Christodoulides},\ and\ \citenamefont
  {Peschel}}]{Regensburger2012}%
  \BibitemOpen
  \bibfield  {author} {\bibinfo {author} {\bibfnamefont {A.}~\bibnamefont
  {Regensburger}}, \bibinfo {author} {\bibfnamefont {C.}~\bibnamefont
  {Bersch}}, \bibinfo {author} {\bibfnamefont {M.-A.}\ \bibnamefont {Miri}},
  \bibinfo {author} {\bibfnamefont {G.}~\bibnamefont {Onishchukov}}, \bibinfo
  {author} {\bibfnamefont {D.~N.}\ \bibnamefont {Christodoulides}},\ and\
  \bibinfo {author} {\bibfnamefont {U.}~\bibnamefont {Peschel}},\ }\bibfield
  {title} {\bibinfo {title} {Parity{\textendash}time synthetic photonic
  lattices},\ }\href {https://doi.org/10.1038/nature11298} {\bibfield
  {journal} {\bibinfo  {journal} {Nature}\ }\textbf {\bibinfo {volume} {488}},\
  \bibinfo {pages} {167} (\bibinfo {year} {2012})}\BibitemShut {NoStop}%
\bibitem [{\citenamefont {Peng}\ \emph {et~al.}(2014)\citenamefont {Peng},
  \citenamefont {\"{O}zdemir}, \citenamefont {Lei}, \citenamefont {Monifi},
  \citenamefont {Gianfreda}, \citenamefont {Long}, \citenamefont {Fan},
  \citenamefont {Nori}, \citenamefont {Bender},\ and\ \citenamefont
  {Yang}}]{Peng2014}%
  \BibitemOpen
  \bibfield  {author} {\bibinfo {author} {\bibfnamefont {B.}~\bibnamefont
  {Peng}}, \bibinfo {author} {\bibfnamefont {{\c{S}}.~K.}\ \bibnamefont
  {\"{O}zdemir}}, \bibinfo {author} {\bibfnamefont {F.}~\bibnamefont {Lei}},
  \bibinfo {author} {\bibfnamefont {F.}~\bibnamefont {Monifi}}, \bibinfo
  {author} {\bibfnamefont {M.}~\bibnamefont {Gianfreda}}, \bibinfo {author}
  {\bibfnamefont {G.~L.}\ \bibnamefont {Long}}, \bibinfo {author}
  {\bibfnamefont {S.}~\bibnamefont {Fan}}, \bibinfo {author} {\bibfnamefont
  {F.}~\bibnamefont {Nori}}, \bibinfo {author} {\bibfnamefont {C.~M.}\
  \bibnamefont {Bender}},\ and\ \bibinfo {author} {\bibfnamefont
  {L.}~\bibnamefont {Yang}},\ }\bibfield  {title} {\bibinfo {title}
  {Parity{\textendash}time-symmetric whispering-gallery microcavities},\ }\href
  {https://doi.org/10.1038/nphys2927} {\bibfield  {journal} {\bibinfo
  {journal} {Nature Physics}\ }\textbf {\bibinfo {volume} {10}},\ \bibinfo
  {pages} {394} (\bibinfo {year} {2014})}\BibitemShut {NoStop}%
\bibitem [{\citenamefont {Hodaei}\ \emph {et~al.}(2017)\citenamefont {Hodaei},
  \citenamefont {Hassan}, \citenamefont {Wittek}, \citenamefont
  {Garcia-Gracia}, \citenamefont {El-Ganainy}, \citenamefont
  {Christodoulides},\ and\ \citenamefont {Khajavikhan}}]{Hodaei2017}%
  \BibitemOpen
  \bibfield  {author} {\bibinfo {author} {\bibfnamefont {H.}~\bibnamefont
  {Hodaei}}, \bibinfo {author} {\bibfnamefont {A.~U.}\ \bibnamefont {Hassan}},
  \bibinfo {author} {\bibfnamefont {S.}~\bibnamefont {Wittek}}, \bibinfo
  {author} {\bibfnamefont {H.}~\bibnamefont {Garcia-Gracia}}, \bibinfo {author}
  {\bibfnamefont {R.}~\bibnamefont {El-Ganainy}}, \bibinfo {author}
  {\bibfnamefont {D.~N.}\ \bibnamefont {Christodoulides}},\ and\ \bibinfo
  {author} {\bibfnamefont {M.}~\bibnamefont {Khajavikhan}},\ }\bibfield
  {title} {\bibinfo {title} {Enhanced sensitivity at higher-order exceptional
  points},\ }\href {https://doi.org/10.1038/nature23280} {\bibfield  {journal}
  {\bibinfo  {journal} {Nature}\ }\textbf {\bibinfo {volume} {548}},\ \bibinfo
  {pages} {187} (\bibinfo {year} {2017})}\BibitemShut {NoStop}%
\bibitem [{\citenamefont {Chen}\ \emph {et~al.}(2017)\citenamefont {Chen},
  \citenamefont {\"{O}zdemir}, \citenamefont {Zhao}, \citenamefont {Wiersig},\
  and\ \citenamefont {Yang}}]{Chen2017}%
  \BibitemOpen
  \bibfield  {author} {\bibinfo {author} {\bibfnamefont {W.}~\bibnamefont
  {Chen}}, \bibinfo {author} {\bibfnamefont {{\c{S}}.~K.}\ \bibnamefont
  {\"{O}zdemir}}, \bibinfo {author} {\bibfnamefont {G.}~\bibnamefont {Zhao}},
  \bibinfo {author} {\bibfnamefont {J.}~\bibnamefont {Wiersig}},\ and\ \bibinfo
  {author} {\bibfnamefont {L.}~\bibnamefont {Yang}},\ }\bibfield  {title}
  {\bibinfo {title} {Exceptional points enhance sensing in an optical
  microcavity},\ }\href {https://doi.org/10.1038/nature23281} {\bibfield
  {journal} {\bibinfo  {journal} {Nature}\ }\textbf {\bibinfo {volume} {548}},\
  \bibinfo {pages} {192} (\bibinfo {year} {2017})}\BibitemShut {NoStop}%
\bibitem [{\citenamefont {Zhu}\ \emph {et~al.}(2014)\citenamefont {Zhu},
  \citenamefont {Ramezani}, \citenamefont {Shi}, \citenamefont {Zhu},\ and\
  \citenamefont {Zhang}}]{Zhu2014}%
  \BibitemOpen
  \bibfield  {author} {\bibinfo {author} {\bibfnamefont {X.}~\bibnamefont
  {Zhu}}, \bibinfo {author} {\bibfnamefont {H.}~\bibnamefont {Ramezani}},
  \bibinfo {author} {\bibfnamefont {C.}~\bibnamefont {Shi}}, \bibinfo {author}
  {\bibfnamefont {J.}~\bibnamefont {Zhu}},\ and\ \bibinfo {author}
  {\bibfnamefont {X.}~\bibnamefont {Zhang}},\ }\bibfield  {title} {\bibinfo
  {title} {$\mathcal{P}\mathcal{T}$-symmetric acoustics},\ }\href
  {https://doi.org/10.1103/PhysRevX.4.031042} {\bibfield  {journal} {\bibinfo
  {journal} {Phys. Rev. X}\ }\textbf {\bibinfo {volume} {4}},\ \bibinfo {pages}
  {031042} (\bibinfo {year} {2014})}\BibitemShut {NoStop}%
\bibitem [{\citenamefont {Li}\ \emph {et~al.}(2019{\natexlab{a}})\citenamefont
  {Li}, \citenamefont {Peng}, \citenamefont {Han}, \citenamefont {Miri},
  \citenamefont {Li}, \citenamefont {Xiao}, \citenamefont {Zhu}, \citenamefont
  {Zhao}, \citenamefont {Al{\`u}}, \citenamefont {Fan},\ and\ \citenamefont
  {Qiu}}]{LiY2019}%
  \BibitemOpen
  \bibfield  {author} {\bibinfo {author} {\bibfnamefont {Y.}~\bibnamefont
  {Li}}, \bibinfo {author} {\bibfnamefont {Y.-G.}\ \bibnamefont {Peng}},
  \bibinfo {author} {\bibfnamefont {L.}~\bibnamefont {Han}}, \bibinfo {author}
  {\bibfnamefont {M.-A.}\ \bibnamefont {Miri}}, \bibinfo {author}
  {\bibfnamefont {W.}~\bibnamefont {Li}}, \bibinfo {author} {\bibfnamefont
  {M.}~\bibnamefont {Xiao}}, \bibinfo {author} {\bibfnamefont {X.-F.}\
  \bibnamefont {Zhu}}, \bibinfo {author} {\bibfnamefont {J.}~\bibnamefont
  {Zhao}}, \bibinfo {author} {\bibfnamefont {A.}~\bibnamefont {Al{\`u}}},
  \bibinfo {author} {\bibfnamefont {S.}~\bibnamefont {Fan}},\ and\ \bibinfo
  {author} {\bibfnamefont {C.-W.}\ \bibnamefont {Qiu}},\ }\bibfield  {title}
  {\bibinfo {title} {Anti{\textendash}parity-time symmetry in diffusive
  systems},\ }\href {https://doi.org/10.1126/science.aaw6259} {\bibfield
  {journal} {\bibinfo  {journal} {Science}\ }\textbf {\bibinfo {volume}
  {364}},\ \bibinfo {pages} {170} (\bibinfo {year}
  {2019}{\natexlab{a}})}\BibitemShut {NoStop}%
\bibitem [{\citenamefont {Humire}\ \emph {et~al.}(2019)\citenamefont {Humire},
  \citenamefont {Joglekar},\ and\ \citenamefont {Nustes}}]{Humire2019}%
  \BibitemOpen
  \bibfield  {author} {\bibinfo {author} {\bibfnamefont {F.~R.}\ \bibnamefont
  {Humire}}, \bibinfo {author} {\bibfnamefont {Y.~N.}\ \bibnamefont
  {Joglekar}},\ and\ \bibinfo {author} {\bibfnamefont {M.~A.~G.}\ \bibnamefont
  {Nustes}},\ }\href@noop {} {\bibinfo {title} {Rabi oscillations of
  dissipative structures effected by out-of-phase parametric drives}} (\bibinfo
  {year} {2019}),\ \Eprint {https://arxiv.org/abs/1905.03896} {arXiv:1905.03896
  [nlin.PS]} \BibitemShut {NoStop}%
\bibitem [{\citenamefont {Vemuri}\ \emph {et~al.}(2021)\citenamefont {Vemuri},
  \citenamefont {Wilkey},\ and\ \citenamefont {Joglekar}}]{Vemuri2021}%
  \BibitemOpen
  \bibfield  {author} {\bibinfo {author} {\bibfnamefont {G.}~\bibnamefont
  {Vemuri}}, \bibinfo {author} {\bibfnamefont {A.}~\bibnamefont {Wilkey}},\
  and\ \bibinfo {author} {\bibfnamefont {Y.~N.}\ \bibnamefont {Joglekar}},\
  }\bibfield  {title} {\bibinfo {title} {Exceptional points in anti-{PT}
  symmetric system of delay coupled semiconductor lasers},\ }in\ \href
  {https://doi.org/10.1117/12.2593311} {\emph {\bibinfo {booktitle} {Active
  Photonic Platforms {XIII}}}},\ \bibinfo {editor} {edited by\ \bibinfo
  {editor} {\bibfnamefont {G.~S.}\ \bibnamefont {Subramania}}\ and\ \bibinfo
  {editor} {\bibfnamefont {S.}~\bibnamefont {Foteinopoulou}}}\ (\bibinfo
  {publisher} {{SPIE}},\ \bibinfo {year} {2021})\BibitemShut {NoStop}%
\bibitem [{\citenamefont {Wu}\ \emph {et~al.}(2019)\citenamefont {Wu},
  \citenamefont {Liu}, \citenamefont {Geng}, \citenamefont {Song},
  \citenamefont {Ye}, \citenamefont {Duan}, \citenamefont {Rong},\ and\
  \citenamefont {Du}}]{Wu2019}%
  \BibitemOpen
  \bibfield  {author} {\bibinfo {author} {\bibfnamefont {Y.}~\bibnamefont
  {Wu}}, \bibinfo {author} {\bibfnamefont {W.}~\bibnamefont {Liu}}, \bibinfo
  {author} {\bibfnamefont {J.}~\bibnamefont {Geng}}, \bibinfo {author}
  {\bibfnamefont {X.}~\bibnamefont {Song}}, \bibinfo {author} {\bibfnamefont
  {X.}~\bibnamefont {Ye}}, \bibinfo {author} {\bibfnamefont {C.-K.}\
  \bibnamefont {Duan}}, \bibinfo {author} {\bibfnamefont {X.}~\bibnamefont
  {Rong}},\ and\ \bibinfo {author} {\bibfnamefont {J.}~\bibnamefont {Du}},\
  }\bibfield  {title} {\bibinfo {title} {Observation of parity-time symmetry
  breaking in a single-spin system},\ }\href
  {https://doi.org/10.1126/science.aaw8205} {\bibfield  {journal} {\bibinfo
  {journal} {Science}\ }\textbf {\bibinfo {volume} {364}},\ \bibinfo {pages}
  {878} (\bibinfo {year} {2019})}\BibitemShut {NoStop}%
\bibitem [{\citenamefont {Naghiloo}\ \emph {et~al.}(2019)\citenamefont
  {Naghiloo}, \citenamefont {Abbasi}, \citenamefont {Joglekar},\ and\
  \citenamefont {Murch}}]{Naghiloo2019}%
  \BibitemOpen
  \bibfield  {author} {\bibinfo {author} {\bibfnamefont {M.}~\bibnamefont
  {Naghiloo}}, \bibinfo {author} {\bibfnamefont {M.}~\bibnamefont {Abbasi}},
  \bibinfo {author} {\bibfnamefont {Y.~N.}\ \bibnamefont {Joglekar}},\ and\
  \bibinfo {author} {\bibfnamefont {K.~W.}\ \bibnamefont {Murch}},\ }\bibfield
  {title} {\bibinfo {title} {Quantum state tomography across the exceptional
  point in a single dissipative qubit},\ }\href
  {https://doi.org/10.1038/s41567-019-0652-z} {\bibfield  {journal} {\bibinfo
  {journal} {Nature Physics}\ }\textbf {\bibinfo {volume} {15}},\ \bibinfo
  {pages} {1232} (\bibinfo {year} {2019})}\BibitemShut {NoStop}%
\bibitem [{\citenamefont {Li}\ \emph {et~al.}(2019{\natexlab{b}})\citenamefont
  {Li}, \citenamefont {Harter}, \citenamefont {Liu}, \citenamefont {de~Melo},
  \citenamefont {Joglekar},\ and\ \citenamefont {Luo}}]{Li2019}%
  \BibitemOpen
  \bibfield  {author} {\bibinfo {author} {\bibfnamefont {J.}~\bibnamefont
  {Li}}, \bibinfo {author} {\bibfnamefont {A.~K.}\ \bibnamefont {Harter}},
  \bibinfo {author} {\bibfnamefont {J.}~\bibnamefont {Liu}}, \bibinfo {author}
  {\bibfnamefont {L.}~\bibnamefont {de~Melo}}, \bibinfo {author} {\bibfnamefont
  {Y.~N.}\ \bibnamefont {Joglekar}},\ and\ \bibinfo {author} {\bibfnamefont
  {L.}~\bibnamefont {Luo}},\ }\bibfield  {title} {\bibinfo {title} {Observation
  of parity-time symmetry breaking transitions in a dissipative floquet system
  of ultracold atoms},\ }\bibfield  {journal} {\bibinfo  {journal} {Nature
  Communications}\ }\textbf {\bibinfo {volume} {10}},\ \href
  {https://doi.org/10.1038/s41467-019-08596-1} {10.1038/s41467-019-08596-1}
  (\bibinfo {year} {2019}{\natexlab{b}})\BibitemShut {NoStop}%
\bibitem [{\citenamefont {Klauck}\ \emph {et~al.}(2019)\citenamefont {Klauck},
  \citenamefont {Teuber}, \citenamefont {Ornigotti}, \citenamefont {Heinrich},
  \citenamefont {Scheel},\ and\ \citenamefont {Szameit}}]{Klauck2019}%
  \BibitemOpen
  \bibfield  {author} {\bibinfo {author} {\bibfnamefont {F.}~\bibnamefont
  {Klauck}}, \bibinfo {author} {\bibfnamefont {L.}~\bibnamefont {Teuber}},
  \bibinfo {author} {\bibfnamefont {M.}~\bibnamefont {Ornigotti}}, \bibinfo
  {author} {\bibfnamefont {M.}~\bibnamefont {Heinrich}}, \bibinfo {author}
  {\bibfnamefont {S.}~\bibnamefont {Scheel}},\ and\ \bibinfo {author}
  {\bibfnamefont {A.}~\bibnamefont {Szameit}},\ }\bibfield  {title} {\bibinfo
  {title} {Observation of $\mathcal{PT}$-symmetric quantum interference},\
  }\href {https://doi.org/10.1038/s41566-019-0517-0} {\bibfield  {journal}
  {\bibinfo  {journal} {Nature Photonics}\ }\textbf {\bibinfo {volume} {13}},\
  \bibinfo {pages} {883} (\bibinfo {year} {2019})}\BibitemShut {NoStop}%
\bibitem [{\citenamefont {Chen}\ \emph {et~al.}(2021)\citenamefont {Chen},
  \citenamefont {Abbasi}, \citenamefont {Joglekar},\ and\ \citenamefont
  {Murch}}]{Chen2021}%
  \BibitemOpen
  \bibfield  {author} {\bibinfo {author} {\bibfnamefont {W.}~\bibnamefont
  {Chen}}, \bibinfo {author} {\bibfnamefont {M.}~\bibnamefont {Abbasi}},
  \bibinfo {author} {\bibfnamefont {Y.~N.}\ \bibnamefont {Joglekar}},\ and\
  \bibinfo {author} {\bibfnamefont {K.~W.}\ \bibnamefont {Murch}},\ }\bibfield
  {title} {\bibinfo {title} {Quantum jumps in the non-hermitian dynamics of a
  superconducting qubit},\ }\bibfield  {journal} {\bibinfo  {journal} {Physical
  Review Letters}\ }\textbf {\bibinfo {volume} {127}},\ \href
  {https://doi.org/10.1103/physrevlett.127.140504}
  {10.1103/physrevlett.127.140504} (\bibinfo {year} {2021})\BibitemShut
  {NoStop}%
\bibitem [{\citenamefont {Caves}(1982)}]{Caves1982}%
  \BibitemOpen
  \bibfield  {author} {\bibinfo {author} {\bibfnamefont {C.~M.}\ \bibnamefont
  {Caves}},\ }\bibfield  {title} {\bibinfo {title} {Quantum limits on noise in
  linear amplifiers},\ }\href {https://doi.org/10.1103/physrevd.26.1817}
  {\bibfield  {journal} {\bibinfo  {journal} {Physical Review D}\ }\textbf
  {\bibinfo {volume} {26}},\ \bibinfo {pages} {1817} (\bibinfo {year}
  {1982})}\BibitemShut {NoStop}%
\bibitem [{\citenamefont {Purkayastha}\ \emph {et~al.}(2020)\citenamefont
  {Purkayastha}, \citenamefont {Kulkarni},\ and\ \citenamefont
  {Joglekar}}]{Purkayastha2020}%
  \BibitemOpen
  \bibfield  {author} {\bibinfo {author} {\bibfnamefont {A.}~\bibnamefont
  {Purkayastha}}, \bibinfo {author} {\bibfnamefont {M.}~\bibnamefont
  {Kulkarni}},\ and\ \bibinfo {author} {\bibfnamefont {Y.~N.}\ \bibnamefont
  {Joglekar}},\ }\bibfield  {title} {\bibinfo {title} {Emergent $\mathcal{PT}$
  symmetry in a double-quantum-dot circuit qed setup},\ }\href
  {https://doi.org/10.1103/PhysRevResearch.2.043075} {\bibfield  {journal}
  {\bibinfo  {journal} {Phys. Rev. Research}\ }\textbf {\bibinfo {volume}
  {2}},\ \bibinfo {pages} {043075} (\bibinfo {year} {2020})}\BibitemShut
  {NoStop}%
\bibitem [{\citenamefont {Scheel}\ and\ \citenamefont
  {Szameit}(2018)}]{Scheel2018}%
  \BibitemOpen
  \bibfield  {author} {\bibinfo {author} {\bibfnamefont {S.}~\bibnamefont
  {Scheel}}\ and\ \bibinfo {author} {\bibfnamefont {A.}~\bibnamefont
  {Szameit}},\ }\bibfield  {title} {\bibinfo {title} {$\mathcal{PT}$-symmetric
  photonic quantum systems with gain and loss do not exist},\ }\href
  {https://doi.org/10.1209/0295-5075/122/34001} {\bibfield  {journal} {\bibinfo
   {journal} {{EPL} (Europhysics Letters)}\ }\textbf {\bibinfo {volume}
  {122}},\ \bibinfo {pages} {34001} (\bibinfo {year} {2018})}\BibitemShut
  {NoStop}%
\bibitem [{\citenamefont {Miri}\ and\ \citenamefont
  {Al{\`{u}}}(2016)}]{Miri2016}%
  \BibitemOpen
  \bibfield  {author} {\bibinfo {author} {\bibfnamefont {M.-A.}\ \bibnamefont
  {Miri}}\ and\ \bibinfo {author} {\bibfnamefont {A.}~\bibnamefont
  {Al{\`{u}}}},\ }\bibfield  {title} {\bibinfo {title} {Nonlinearity-induced
  {PT}-symmetry without material gain},\ }\href
  {https://doi.org/10.1088/1367-2630/18/6/065001} {\bibfield  {journal}
  {\bibinfo  {journal} {New Journal of Physics}\ }\textbf {\bibinfo {volume}
  {18}},\ \bibinfo {pages} {065001} (\bibinfo {year} {2016})}\BibitemShut
  {NoStop}%
\bibitem [{\citenamefont {Roy}\ \emph {et~al.}(2021)\citenamefont {Roy},
  \citenamefont {Jahani}, \citenamefont {Guo}, \citenamefont {Dutt},
  \citenamefont {Fan}, \citenamefont {Miri},\ and\ \citenamefont
  {Marandi}}]{Roy2021}%
  \BibitemOpen
  \bibfield  {author} {\bibinfo {author} {\bibfnamefont {A.}~\bibnamefont
  {Roy}}, \bibinfo {author} {\bibfnamefont {S.}~\bibnamefont {Jahani}},
  \bibinfo {author} {\bibfnamefont {Q.}~\bibnamefont {Guo}}, \bibinfo {author}
  {\bibfnamefont {A.}~\bibnamefont {Dutt}}, \bibinfo {author} {\bibfnamefont
  {S.}~\bibnamefont {Fan}}, \bibinfo {author} {\bibfnamefont {M.-A.}\
  \bibnamefont {Miri}},\ and\ \bibinfo {author} {\bibfnamefont
  {A.}~\bibnamefont {Marandi}},\ }\bibfield  {title} {\bibinfo {title}
  {Nondissipative non-hermitian dynamics and exceptional points in coupled
  optical parametric oscillators},\ }\href
  {https://doi.org/10.1364/optica.415569} {\bibfield  {journal} {\bibinfo
  {journal} {Optica}\ }\textbf {\bibinfo {volume} {8}},\ \bibinfo {pages} {415}
  (\bibinfo {year} {2021})}\BibitemShut {NoStop}%
\bibitem [{\citenamefont {Li}\ \emph {et~al.}(2020)\citenamefont {Li},
  \citenamefont {Moussa}, \citenamefont {Sounas},\ and\ \citenamefont
  {Al\`u}}]{Hunan2020}%
  \BibitemOpen
  \bibfield  {author} {\bibinfo {author} {\bibfnamefont {H.}~\bibnamefont
  {Li}}, \bibinfo {author} {\bibfnamefont {H.}~\bibnamefont {Moussa}}, \bibinfo
  {author} {\bibfnamefont {D.}~\bibnamefont {Sounas}},\ and\ \bibinfo {author}
  {\bibfnamefont {A.}~\bibnamefont {Al\`u}},\ }\bibfield  {title} {\bibinfo
  {title} {Parity-time symmetry based on time modulation},\ }\href
  {https://doi.org/10.1103/PhysRevApplied.14.031002} {\bibfield  {journal}
  {\bibinfo  {journal} {Phys. Rev. Applied}\ }\textbf {\bibinfo {volume}
  {14}},\ \bibinfo {pages} {031002} (\bibinfo {year} {2020})}\BibitemShut
  {NoStop}%
\bibitem [{\citenamefont {de~J.~Le{\'{o}}n-Montiel}\ \emph
  {et~al.}(2018)\citenamefont {de~J.~Le{\'{o}}n-Montiel}, \citenamefont
  {Quiroz-Ju{\'{a}}rez}, \citenamefont {Dom{\'{\i}}nguez-Ju{\'{a}}rez},
  \citenamefont {Quintero-Torres}, \citenamefont {Arag{\'{o}}n}, \citenamefont
  {Harter},\ and\ \citenamefont {Joglekar}}]{LeonMontiel2018}%
  \BibitemOpen
  \bibfield  {author} {\bibinfo {author} {\bibfnamefont {R.}~\bibnamefont
  {de~J.~Le{\'{o}}n-Montiel}}, \bibinfo {author} {\bibfnamefont {M.~A.}\
  \bibnamefont {Quiroz-Ju{\'{a}}rez}}, \bibinfo {author} {\bibfnamefont
  {J.~L.}\ \bibnamefont {Dom{\'{\i}}nguez-Ju{\'{a}}rez}}, \bibinfo {author}
  {\bibfnamefont {R.}~\bibnamefont {Quintero-Torres}}, \bibinfo {author}
  {\bibfnamefont {J.~L.}\ \bibnamefont {Arag{\'{o}}n}}, \bibinfo {author}
  {\bibfnamefont {A.~K.}\ \bibnamefont {Harter}},\ and\ \bibinfo {author}
  {\bibfnamefont {Y.~N.}\ \bibnamefont {Joglekar}},\ }\bibfield  {title}
  {\bibinfo {title} {Observation of slowly decaying eigenmodes without
  exceptional points in floquet dissipative synthetic circuits},\ }\bibfield
  {journal} {\bibinfo  {journal} {Communications Physics}\ }\textbf {\bibinfo
  {volume} {1}},\ \href {https://doi.org/10.1038/s42005-018-0087-3}
  {10.1038/s42005-018-0087-3} (\bibinfo {year} {2018})\BibitemShut {NoStop}%
\bibitem [{\citenamefont {Quiroz-Ju\'arez}\ \emph {et~al.}(2021)\citenamefont
  {Quiroz-Ju\'arez}, \citenamefont {You}, \citenamefont
  {Carrillo-Mart\'{\i}nez}, \citenamefont {Montiel-\'Alvarez}, \citenamefont
  {Arag\'on}, \citenamefont {Maga\~na Loaiza},\ and\ \citenamefont
  {de~J.~Le{\'{o}}n-Montiel}}]{Alan2021}%
  \BibitemOpen
  \bibfield  {author} {\bibinfo {author} {\bibfnamefont {M.~A.}\ \bibnamefont
  {Quiroz-Ju\'arez}}, \bibinfo {author} {\bibfnamefont {C.}~\bibnamefont
  {You}}, \bibinfo {author} {\bibfnamefont {J.}~\bibnamefont
  {Carrillo-Mart\'{\i}nez}}, \bibinfo {author} {\bibfnamefont {D.}~\bibnamefont
  {Montiel-\'Alvarez}}, \bibinfo {author} {\bibfnamefont {J.~L.}\ \bibnamefont
  {Arag\'on}}, \bibinfo {author} {\bibfnamefont {O.~S.}\ \bibnamefont {Maga\~na
  Loaiza}},\ and\ \bibinfo {author} {\bibfnamefont {R.}~\bibnamefont
  {de~J.~Le{\'{o}}n-Montiel}},\ }\bibfield  {title} {\bibinfo {title}
  {Reconfigurable network for quantum transport simulations},\ }\href
  {https://doi.org/10.1103/PhysRevResearch.3.013010} {\bibfield  {journal}
  {\bibinfo  {journal} {Phys. Rev. Research}\ }\textbf {\bibinfo {volume}
  {3}},\ \bibinfo {pages} {013010} (\bibinfo {year} {2021})}\BibitemShut
  {NoStop}%
\bibitem [{\citenamefont {Nakahara}(2003)}]{Nakahara2003}%
  \BibitemOpen
  \bibfield  {author} {\bibinfo {author} {\bibfnamefont {M.}~\bibnamefont
  {Nakahara}},\ }\href@noop {} {\emph {\bibinfo {title} {Geometry, topology,
  and physics}}}\ (\bibinfo  {publisher} {Institute of Physics Pub. CRC Press,
  an imprint of Taylor \& Francis Group},\ \bibinfo {address} {Bristol
  Philadelphia Boca Raton Fla},\ \bibinfo {year} {2003})\BibitemShut {NoStop}%
\bibitem [{\citenamefont {Quiroz-Ju\'arez}\ \emph {et~al.}(2020)\citenamefont
  {Quiroz-Ju\'arez}, \citenamefont {Ch\'avez-Carlos}, \citenamefont {Arag\'on},
  \citenamefont {Hirsch},\ and\ \citenamefont
  {de~J.~Le{\'{o}}n-Montiel}}]{Alan2020}%
  \BibitemOpen
  \bibfield  {author} {\bibinfo {author} {\bibfnamefont {M.~A.}\ \bibnamefont
  {Quiroz-Ju\'arez}}, \bibinfo {author} {\bibfnamefont {J.}~\bibnamefont
  {Ch\'avez-Carlos}}, \bibinfo {author} {\bibfnamefont {J.~L.}\ \bibnamefont
  {Arag\'on}}, \bibinfo {author} {\bibfnamefont {J.~G.}\ \bibnamefont
  {Hirsch}},\ and\ \bibinfo {author} {\bibfnamefont {R.}~\bibnamefont
  {de~J.~Le{\'{o}}n-Montiel}},\ }\bibfield  {title} {\bibinfo {title}
  {Experimental realization of the classical dicke model},\ }\href
  {https://doi.org/10.1103/PhysRevResearch.2.033169} {\bibfield  {journal}
  {\bibinfo  {journal} {Phys. Rev. Research}\ }\textbf {\bibinfo {volume}
  {2}},\ \bibinfo {pages} {033169} (\bibinfo {year} {2020})}\BibitemShut
  {NoStop}%
\bibitem [{\citenamefont {Bian}\ \emph {et~al.}(2020)\citenamefont {Bian},
  \citenamefont {Xiao}, \citenamefont {Wang}, \citenamefont {Zhan},
  \citenamefont {Onanga}, \citenamefont {Ruzicka}, \citenamefont {Yi},
  \citenamefont {Joglekar},\ and\ \citenamefont {Xue}}]{Bian2020}%
  \BibitemOpen
  \bibfield  {author} {\bibinfo {author} {\bibfnamefont {Z.}~\bibnamefont
  {Bian}}, \bibinfo {author} {\bibfnamefont {L.}~\bibnamefont {Xiao}}, \bibinfo
  {author} {\bibfnamefont {K.}~\bibnamefont {Wang}}, \bibinfo {author}
  {\bibfnamefont {X.}~\bibnamefont {Zhan}}, \bibinfo {author} {\bibfnamefont
  {F.~A.}\ \bibnamefont {Onanga}}, \bibinfo {author} {\bibfnamefont
  {F.}~\bibnamefont {Ruzicka}}, \bibinfo {author} {\bibfnamefont
  {W.}~\bibnamefont {Yi}}, \bibinfo {author} {\bibfnamefont {Y.~N.}\
  \bibnamefont {Joglekar}},\ and\ \bibinfo {author} {\bibfnamefont
  {P.}~\bibnamefont {Xue}},\ }\bibfield  {title} {\bibinfo {title} {Conserved
  quantities in parity-time symmetric systems},\ }\bibfield  {journal}
  {\bibinfo  {journal} {Physical Review Research}\ }\textbf {\bibinfo {volume}
  {2}},\ \href {https://doi.org/10.1103/physrevresearch.2.022039}
  {10.1103/physrevresearch.2.022039} (\bibinfo {year} {2020})\BibitemShut
  {NoStop}%
\bibitem [{\citenamefont {Ruzicka}\ \emph {et~al.}(2021)\citenamefont
  {Ruzicka}, \citenamefont {Agarwal},\ and\ \citenamefont
  {Joglekar}}]{Ruzicka2021}%
  \BibitemOpen
  \bibfield  {author} {\bibinfo {author} {\bibfnamefont {F.}~\bibnamefont
  {Ruzicka}}, \bibinfo {author} {\bibfnamefont {K.~S.}\ \bibnamefont
  {Agarwal}},\ and\ \bibinfo {author} {\bibfnamefont {Y.~N.}\ \bibnamefont
  {Joglekar}},\ }\href@noop {} {\bibinfo {title} {Conserved quantities,
  exceptional points, and antilinear symmetries in non-hermitian systems}}
  (\bibinfo {year} {2021}),\ \Eprint {https://arxiv.org/abs/2104.11265}
  {arXiv:2104.11265 [quant-ph]} \BibitemShut {NoStop}%
\bibitem [{\citenamefont {Mostafazadeh}(2002)}]{Mostafazadeh2002}%
  \BibitemOpen
  \bibfield  {author} {\bibinfo {author} {\bibfnamefont {A.}~\bibnamefont
  {Mostafazadeh}},\ }\bibfield  {title} {\bibinfo {title} {Pseudo-hermiticity
  versus {PT} symmetry: The necessary condition for the reality of the spectrum
  of a non-hermitian hamiltonian},\ }\href {https://doi.org/10.1063/1.1418246}
  {\bibfield  {journal} {\bibinfo  {journal} {Journal of Mathematical Physics}\
  }\textbf {\bibinfo {volume} {43}},\ \bibinfo {pages} {205} (\bibinfo {year}
  {2002})}\BibitemShut {NoStop}%
\bibitem [{\citenamefont {Mostafazadeh}(2010)}]{Mostafazadeh2010}%
  \BibitemOpen
  \bibfield  {author} {\bibinfo {author} {\bibfnamefont {A.}~\bibnamefont
  {Mostafazadeh}},\ }\bibfield  {title} {\bibinfo {title} {Pseudo-hermitian
  representation of quantum mechanics},\ }\href
  {https://doi.org/10.1142/s0219887810004816} {\bibfield  {journal} {\bibinfo
  {journal} {International Journal of Geometric Methods in Modern Physics}\
  }\textbf {\bibinfo {volume} {07}},\ \bibinfo {pages} {1191} (\bibinfo {year}
  {2010})}\BibitemShut {NoStop}%
\bibitem [{\citenamefont {Hanggi}(2020)}]{Hanggi1998}%
  \BibitemOpen
  \bibfield  {author} {\bibinfo {author} {\bibfnamefont {P.}~\bibnamefont
  {Hanggi}},\ }\href
  {ttps://www.physik.uni-augsburg.de/theo1/hanggi/Chapter_5.pdf} {\emph
  {\bibinfo {title} {Diven quantum systems}}} (\bibinfo {year} {1998 (accessed
  October 31, 2020)})\BibitemShut {NoStop}%
\bibitem [{\citenamefont {Luo}\ \emph {et~al.}(2013)\citenamefont {Luo},
  \citenamefont {Huang}, \citenamefont {Zhong}, \citenamefont {Qin},
  \citenamefont {Xie}, \citenamefont {Kivshar},\ and\ \citenamefont
  {Lee}}]{Luo2013}%
  \BibitemOpen
  \bibfield  {author} {\bibinfo {author} {\bibfnamefont {X.}~\bibnamefont
  {Luo}}, \bibinfo {author} {\bibfnamefont {J.}~\bibnamefont {Huang}}, \bibinfo
  {author} {\bibfnamefont {H.}~\bibnamefont {Zhong}}, \bibinfo {author}
  {\bibfnamefont {X.}~\bibnamefont {Qin}}, \bibinfo {author} {\bibfnamefont
  {Q.}~\bibnamefont {Xie}}, \bibinfo {author} {\bibfnamefont {Y.~S.}\
  \bibnamefont {Kivshar}},\ and\ \bibinfo {author} {\bibfnamefont
  {C.}~\bibnamefont {Lee}},\ }\bibfield  {title} {\bibinfo {title}
  {Pseudo-parity-time symmetry in optical systems},\ }\href
  {https://doi.org/10.1103/PhysRevLett.110.243902} {\bibfield  {journal}
  {\bibinfo  {journal} {Phys. Rev. Lett.}\ }\textbf {\bibinfo {volume} {110}},\
  \bibinfo {pages} {243902} (\bibinfo {year} {2013})}\BibitemShut {NoStop}%
\bibitem [{\citenamefont {Joglekar}\ \emph {et~al.}(2014)\citenamefont
  {Joglekar}, \citenamefont {Marathe}, \citenamefont {Durganandini},\ and\
  \citenamefont {Pathak}}]{Joglekar2014}%
  \BibitemOpen
  \bibfield  {author} {\bibinfo {author} {\bibfnamefont {Y.~N.}\ \bibnamefont
  {Joglekar}}, \bibinfo {author} {\bibfnamefont {R.}~\bibnamefont {Marathe}},
  \bibinfo {author} {\bibfnamefont {P.}~\bibnamefont {Durganandini}},\ and\
  \bibinfo {author} {\bibfnamefont {R.~K.}\ \bibnamefont {Pathak}},\ }\bibfield
   {title} {\bibinfo {title} {{PTspectroscopy} of the rabi problem},\ }\href
  {https://doi.org/10.1103/physreva.90.040101} {\bibfield  {journal} {\bibinfo
  {journal} {Physical Review A}\ }\textbf {\bibinfo {volume} {90}},\ \bibinfo
  {pages} {040101} (\bibinfo {year} {2014})}\BibitemShut {NoStop}%
\bibitem [{\citenamefont {Lee}\ and\ \citenamefont {Joglekar}(2015)}]{Lee2015}%
  \BibitemOpen
  \bibfield  {author} {\bibinfo {author} {\bibfnamefont {T.~E.}\ \bibnamefont
  {Lee}}\ and\ \bibinfo {author} {\bibfnamefont {Y.~N.}\ \bibnamefont
  {Joglekar}},\ }\bibfield  {title} {\bibinfo {title} {$\mathcal{PT}$-symmetric
  rabi model: Perturbation theory},\ }\href
  {https://doi.org/10.1103/PhysRevA.92.042103} {\bibfield  {journal} {\bibinfo
  {journal} {Phys. Rev. A}\ }\textbf {\bibinfo {volume} {92}},\ \bibinfo
  {pages} {042103} (\bibinfo {year} {2015})}\BibitemShut {NoStop}%
\bibitem [{\citenamefont {Chitsazi}\ \emph {et~al.}(2017)\citenamefont
  {Chitsazi}, \citenamefont {Li}, \citenamefont {Ellis},\ and\ \citenamefont
  {Kottos}}]{Chitsazi2017}%
  \BibitemOpen
  \bibfield  {author} {\bibinfo {author} {\bibfnamefont {M.}~\bibnamefont
  {Chitsazi}}, \bibinfo {author} {\bibfnamefont {H.}~\bibnamefont {Li}},
  \bibinfo {author} {\bibfnamefont {F.}~\bibnamefont {Ellis}},\ and\ \bibinfo
  {author} {\bibfnamefont {T.}~\bibnamefont {Kottos}},\ }\bibfield  {title}
  {\bibinfo {title} {Experimental realization of floquet
  $\mathcal{PT}$-symmetric systems},\ }\href
  {https://doi.org/10.1103/physrevlett.119.093901} {\bibfield  {journal}
  {\bibinfo  {journal} {Physical Review Letters}\ }\textbf {\bibinfo {volume}
  {119}},\ \bibinfo {pages} {093901} (\bibinfo {year} {2017})}\BibitemShut
  {NoStop}%
\bibitem [{\citenamefont {Harter}\ and\ \citenamefont
  {Joglekar}(2020)}]{Harter2020}%
  \BibitemOpen
  \bibfield  {author} {\bibinfo {author} {\bibfnamefont {A.~K.}\ \bibnamefont
  {Harter}}\ and\ \bibinfo {author} {\bibfnamefont {Y.~N.}\ \bibnamefont
  {Joglekar}},\ }\bibfield  {title} {\bibinfo {title} {Connecting active and
  passive {PT}-symmetric {F}loquet modulation models},\ }\bibfield  {journal}
  {\bibinfo  {journal} {Progress of Theoretical and Experimental Physics}\
  }\textbf {\bibinfo {volume} {2020}},\ \href
  {https://doi.org/10.1093/ptep/ptaa181} {10.1093/ptep/ptaa181} (\bibinfo
  {year} {2020})\BibitemShut {NoStop}%
\bibitem [{\citenamefont {Kumar}\ \emph {et~al.}(2021)\citenamefont {Kumar},
  \citenamefont {Murch},\ and\ \citenamefont {Joglekar}}]{Akhil2021}%
  \BibitemOpen
  \bibfield  {author} {\bibinfo {author} {\bibfnamefont {A.}~\bibnamefont
  {Kumar}}, \bibinfo {author} {\bibfnamefont {K.~W.}\ \bibnamefont {Murch}},\
  and\ \bibinfo {author} {\bibfnamefont {Y.~N.}\ \bibnamefont {Joglekar}},\
  }\href@noop {} {\bibinfo {title} {Maximal quantum entanglement at exceptional
  points via unitary and thermal dynamics}} (\bibinfo {year} {2021}),\ \Eprint
  {https://arxiv.org/abs/2109.07503} {arXiv:2109.07503 [quant-ph]} \BibitemShut
  {NoStop}%
\bibitem [{\citenamefont {Cochran}\ \emph {et~al.}(2021)\citenamefont
  {Cochran}, \citenamefont {Saxena},\ and\ \citenamefont
  {Joglekar}}]{Cochran2021}%
  \BibitemOpen
  \bibfield  {author} {\bibinfo {author} {\bibfnamefont {Z.~A.}\ \bibnamefont
  {Cochran}}, \bibinfo {author} {\bibfnamefont {A.}~\bibnamefont {Saxena}},\
  and\ \bibinfo {author} {\bibfnamefont {Y.~N.}\ \bibnamefont {Joglekar}},\
  }\bibfield  {title} {\bibinfo {title} {Parity-time symmetric systems with
  memory},\ }\href {https://doi.org/10.1103/PhysRevResearch.3.013135}
  {\bibfield  {journal} {\bibinfo  {journal} {Phys. Rev. Research}\ }\textbf
  {\bibinfo {volume} {3}},\ \bibinfo {pages} {013135} (\bibinfo {year}
  {2021})}\BibitemShut {NoStop}%
\bibitem [{\citenamefont {Doppler}\ \emph {et~al.}(2016)\citenamefont
  {Doppler}, \citenamefont {Mailybaev}, \citenamefont {B\"{o}hm}, \citenamefont
  {Kuhl}, \citenamefont {Girschik}, \citenamefont {Libisch}, \citenamefont
  {Milburn}, \citenamefont {Rabl}, \citenamefont {Moiseyev},\ and\
  \citenamefont {Rotter}}]{Doppler2016}%
  \BibitemOpen
  \bibfield  {author} {\bibinfo {author} {\bibfnamefont {J.}~\bibnamefont
  {Doppler}}, \bibinfo {author} {\bibfnamefont {A.~A.}\ \bibnamefont
  {Mailybaev}}, \bibinfo {author} {\bibfnamefont {J.}~\bibnamefont {B\"{o}hm}},
  \bibinfo {author} {\bibfnamefont {U.}~\bibnamefont {Kuhl}}, \bibinfo {author}
  {\bibfnamefont {A.}~\bibnamefont {Girschik}}, \bibinfo {author}
  {\bibfnamefont {F.}~\bibnamefont {Libisch}}, \bibinfo {author} {\bibfnamefont
  {T.~J.}\ \bibnamefont {Milburn}}, \bibinfo {author} {\bibfnamefont
  {P.}~\bibnamefont {Rabl}}, \bibinfo {author} {\bibfnamefont {N.}~\bibnamefont
  {Moiseyev}},\ and\ \bibinfo {author} {\bibfnamefont {S.}~\bibnamefont
  {Rotter}},\ }\bibfield  {title} {\bibinfo {title} {Dynamically encircling an
  exceptional point for asymmetric mode switching},\ }\href
  {https://doi.org/10.1038/nature18605} {\bibfield  {journal} {\bibinfo
  {journal} {Nature}\ }\textbf {\bibinfo {volume} {537}},\ \bibinfo {pages}
  {76} (\bibinfo {year} {2016})}\BibitemShut {NoStop}%
\bibitem [{\citenamefont {Xu}\ \emph {et~al.}(2016)\citenamefont {Xu},
  \citenamefont {Mason}, \citenamefont {Jiang},\ and\ \citenamefont
  {Harris}}]{Xu2016}%
  \BibitemOpen
  \bibfield  {author} {\bibinfo {author} {\bibfnamefont {H.}~\bibnamefont
  {Xu}}, \bibinfo {author} {\bibfnamefont {D.}~\bibnamefont {Mason}}, \bibinfo
  {author} {\bibfnamefont {L.}~\bibnamefont {Jiang}},\ and\ \bibinfo {author}
  {\bibfnamefont {J.~G.~E.}\ \bibnamefont {Harris}},\ }\bibfield  {title}
  {\bibinfo {title} {Topological energy transfer in an optomechanical system
  with exceptional points},\ }\href {https://doi.org/10.1038/nature18604}
  {\bibfield  {journal} {\bibinfo  {journal} {Nature}\ }\textbf {\bibinfo
  {volume} {537}},\ \bibinfo {pages} {80} (\bibinfo {year} {2016})}\BibitemShut
  {NoStop}%
\bibitem [{\citenamefont {Wang}\ \emph {et~al.}(2021)\citenamefont {Wang},
  \citenamefont {Dutt}, \citenamefont {Wojcik},\ and\ \citenamefont
  {Fan}}]{KWang2021}%
  \BibitemOpen
  \bibfield  {author} {\bibinfo {author} {\bibfnamefont {K.}~\bibnamefont
  {Wang}}, \bibinfo {author} {\bibfnamefont {A.}~\bibnamefont {Dutt}}, \bibinfo
  {author} {\bibfnamefont {C.~C.}\ \bibnamefont {Wojcik}},\ and\ \bibinfo
  {author} {\bibfnamefont {S.}~\bibnamefont {Fan}},\ }\bibfield  {title}
  {\bibinfo {title} {Topological complex-energy braiding of non-hermitian
  bands},\ }\href {https://doi.org/10.1038/s41586-021-03848-x} {\bibfield
  {journal} {\bibinfo  {journal} {Nature}\ }\textbf {\bibinfo {volume} {598}},\
  \bibinfo {pages} {59} (\bibinfo {year} {2021})}\BibitemShut {NoStop}%
\bibitem [{\citenamefont {Bateman}(1931)}]{Bateman1931}%
  \BibitemOpen
  \bibfield  {author} {\bibinfo {author} {\bibfnamefont {H.}~\bibnamefont
  {Bateman}},\ }\bibfield  {title} {\bibinfo {title} {On dissipative systems
  and related variational principles},\ }\href
  {https://doi.org/10.1103/PhysRev.38.815} {\bibfield  {journal} {\bibinfo
  {journal} {Phys. Rev.}\ }\textbf {\bibinfo {volume} {38}},\ \bibinfo {pages}
  {815} (\bibinfo {year} {1931})}\BibitemShut {NoStop}%
\bibitem [{\citenamefont {Kanai}(1948)}]{Kanai1948}%
  \BibitemOpen
  \bibfield  {author} {\bibinfo {author} {\bibfnamefont {E.}~\bibnamefont
  {Kanai}},\ }\bibfield  {title} {\bibinfo {title} {On the quantization of the
  dissipative systems},\ }\href {https://doi.org/10.1143/ptp/3.4.440}
  {\bibfield  {journal} {\bibinfo  {journal} {Progress of Theoretical Physics}\
  }\textbf {\bibinfo {volume} {3}},\ \bibinfo {pages} {440} (\bibinfo {year}
  {1948})}\BibitemShut {NoStop}%
\bibitem [{\citenamefont {Brittin}(1950)}]{Britten1950}%
  \BibitemOpen
  \bibfield  {author} {\bibinfo {author} {\bibfnamefont {W.~E.}\ \bibnamefont
  {Brittin}},\ }\bibfield  {title} {\bibinfo {title} {A note on the
  quantization of dissipative systems},\ }\href
  {https://doi.org/10.1103/PhysRev.77.396} {\bibfield  {journal} {\bibinfo
  {journal} {Phys. Rev.}\ }\textbf {\bibinfo {volume} {77}},\ \bibinfo {pages}
  {396} (\bibinfo {year} {1950})}\BibitemShut {NoStop}%
\bibitem [{\citenamefont {Feshbach}\ and\ \citenamefont
  {Tikochinsky}(1977)}]{Feshbach1977}%
  \BibitemOpen
  \bibfield  {author} {\bibinfo {author} {\bibfnamefont {H.}~\bibnamefont
  {Feshbach}}\ and\ \bibinfo {author} {\bibfnamefont {Y.}~\bibnamefont
  {Tikochinsky}},\ }\bibfield  {title} {\bibinfo {title} {Quantization of the
  damped harmonic oscillator},\ }\href
  {https://doi.org/10.1111/j.2164-0947.1977.tb02946.x} {\bibfield  {journal}
  {\bibinfo  {journal} {Trans. N.Y. Acad. Sci.}\ }\textbf {\bibinfo {volume}
  {38}},\ \bibinfo {pages} {44} (\bibinfo {year} {1977})}\BibitemShut {NoStop}%
\bibitem [{\citenamefont {Greenberger}(1979{\natexlab{a}})}]{Green1979a}%
  \BibitemOpen
  \bibfield  {author} {\bibinfo {author} {\bibfnamefont {D.~M.}\ \bibnamefont
  {Greenberger}},\ }\bibfield  {title} {\bibinfo {title} {A critique of the
  major approaches to damping in quantum theory},\ }\href
  {https://doi.org/10.1063/1.524148} {\bibfield  {journal} {\bibinfo  {journal}
  {J. Math. Phys.}\ }\textbf {\bibinfo {volume} {20}},\ \bibinfo {pages} {762}
  (\bibinfo {year} {1979}{\natexlab{a}})}\BibitemShut {NoStop}%
\bibitem [{\citenamefont {Greenberger}(1979{\natexlab{b}})}]{Green1979b}%
  \BibitemOpen
  \bibfield  {author} {\bibinfo {author} {\bibfnamefont {D.~M.}\ \bibnamefont
  {Greenberger}},\ }\bibfield  {title} {\bibinfo {title} {A new approach to the
  problem of dissipation in quantum mechanics},\ }\href
  {https://doi.org/10.1063/1.524149} {\bibfield  {journal} {\bibinfo  {journal}
  {J. Math. Phys.}\ }\textbf {\bibinfo {volume} {20}},\ \bibinfo {pages} {771}
  (\bibinfo {year} {1979}{\natexlab{b}})}\BibitemShut {NoStop}%
\bibitem [{\citenamefont {Dekker}(1981)}]{Dekker1981}%
  \BibitemOpen
  \bibfield  {author} {\bibinfo {author} {\bibfnamefont {H.}~\bibnamefont
  {Dekker}},\ }\bibfield  {title} {\bibinfo {title} {Classical and quantum
  mechanics of the damped harmonic oscillator},\ }\href
  {https://doi.org/10.1016/0370-1573(81)90033-8} {\bibfield  {journal}
  {\bibinfo  {journal} {Phys. Rep.}\ }\textbf {\bibinfo {volume} {80}},\
  \bibinfo {pages} {1} (\bibinfo {year} {1981})}\BibitemShut {NoStop}%
\bibitem [{\citenamefont {Cheng}\ and\ \citenamefont {Fung}(1988)}]{Cheng1988}%
  \BibitemOpen
  \bibfield  {author} {\bibinfo {author} {\bibfnamefont {C.~M.}\ \bibnamefont
  {Cheng}}\ and\ \bibinfo {author} {\bibfnamefont {P.~C.~W.}\ \bibnamefont
  {Fung}},\ }\bibfield  {title} {\bibinfo {title} {The evolution operator
  technique in solving the schrodinger equation, and its application to
  disentangling exponential operators and solving the problem of a mass-varying
  harmonic oscillator},\ }\href {https://doi.org/10.1088/0305-4470/21/22/015}
  {\bibfield  {journal} {\bibinfo  {journal} {J. Phys. A: Math. Gen.}\ }\textbf
  {\bibinfo {volume} {21}},\ \bibinfo {pages} {4115} (\bibinfo {year}
  {1988})}\BibitemShut {NoStop}%
\bibitem [{\citenamefont {G{\'{o}}mez}(2007)}]{Gomez2007}%
  \BibitemOpen
  \bibfield  {author} {\bibinfo {author} {\bibfnamefont {F.}~\bibnamefont
  {G{\'{o}}mez}},\ }\bibfield  {title} {\bibinfo {title} {Nonunitary similarity
  transformation of conservative to dissipative evolutions: Intertwining
  without time operator},\ }\href {https://doi.org/10.1063/1.2709634}
  {\bibfield  {journal} {\bibinfo  {journal} {J. Math. Phys.}\ }\textbf
  {\bibinfo {volume} {48}},\ \bibinfo {pages} {043506} (\bibinfo {year}
  {2007})}\BibitemShut {NoStop}%
\bibitem [{\citenamefont {Droulias}\ \emph {et~al.}(2019)\citenamefont
  {Droulias}, \citenamefont {Katsantonis}, \citenamefont {Kafesaki},
  \citenamefont {Soukoulis},\ and\ \citenamefont {Economou}}]{Sotiris2019}%
  \BibitemOpen
  \bibfield  {author} {\bibinfo {author} {\bibfnamefont {S.}~\bibnamefont
  {Droulias}}, \bibinfo {author} {\bibfnamefont {I.}~\bibnamefont
  {Katsantonis}}, \bibinfo {author} {\bibfnamefont {M.}~\bibnamefont
  {Kafesaki}}, \bibinfo {author} {\bibfnamefont {C.~M.}\ \bibnamefont
  {Soukoulis}},\ and\ \bibinfo {author} {\bibfnamefont {E.~N.}\ \bibnamefont
  {Economou}},\ }\bibfield  {title} {\bibinfo {title} {Chiral metamaterials
  with $pt$ symmetry and beyond},\ }\href
  {https://doi.org/10.1103/PhysRevLett.122.213201} {\bibfield  {journal}
  {\bibinfo  {journal} {Phys. Rev. Lett.}\ }\textbf {\bibinfo {volume} {122}},\
  \bibinfo {pages} {213201} (\bibinfo {year} {2019})}\BibitemShut {NoStop}%
\bibitem [{\citenamefont {Lee}\ \emph {et~al.}(2018)\citenamefont {Lee},
  \citenamefont {Imhof}, \citenamefont {Berger}, \citenamefont {Bayer},
  \citenamefont {Brehm}, \citenamefont {Molenkamp}, \citenamefont {Kiessling},\
  and\ \citenamefont {Thomale}}]{Lee2018}%
  \BibitemOpen
  \bibfield  {author} {\bibinfo {author} {\bibfnamefont {C.~H.}\ \bibnamefont
  {Lee}}, \bibinfo {author} {\bibfnamefont {S.}~\bibnamefont {Imhof}}, \bibinfo
  {author} {\bibfnamefont {C.}~\bibnamefont {Berger}}, \bibinfo {author}
  {\bibfnamefont {F.}~\bibnamefont {Bayer}}, \bibinfo {author} {\bibfnamefont
  {J.}~\bibnamefont {Brehm}}, \bibinfo {author} {\bibfnamefont {L.~W.}\
  \bibnamefont {Molenkamp}}, \bibinfo {author} {\bibfnamefont {T.}~\bibnamefont
  {Kiessling}},\ and\ \bibinfo {author} {\bibfnamefont {R.}~\bibnamefont
  {Thomale}},\ }\bibfield  {title} {\bibinfo {title} {Topolectrical circuits},\
  }\bibfield  {journal} {\bibinfo  {journal} {Commun Phys}\ }\textbf {\bibinfo
  {volume} {1}},\ \href {https://doi.org/10.1038/s42005-018-0035-2}
  {10.1038/s42005-018-0035-2} (\bibinfo {year} {2018})\BibitemShut {NoStop}%
\end{thebibliography}%



\end{document}